\documentclass[reprint,superscriptaddress,twocolumn,amsmath,amssymb,amsthm, dsfont,prb]{revtex4-1}
\usepackage{graphicx}
\usepackage{mathrsfs}
\usepackage{dcolumn}
\usepackage{bm}
\usepackage{hyperref}
\usepackage{tikz}
\usepackage{hhline}
\usepackage{float}
\usepackage{scalerel}
\usepackage{amsfonts}
\usepackage{bbm}
\usepackage{color}
\usepackage{tabularx}
\usepackage{multirow}
\definecolor{darkblue}{rgb}{0.,0.,0.4}
\definecolor{darkred}{rgb}{0.5,0.,0.}

\usetikzlibrary{decorations.markings}

\makeatletter
\def\l@subsection#1#2{}
\def\l@subsubsection#1#2{}
\makeatother

\begin{document}
\tikzset{
    vertex/.style={fill,circle,draw,scale=0.3},
    ->/.style={decoration={markings,mark=at position 0.5 with {\fill (2pt,0)--(-2pt,2.31pt)--(-2pt,-2.31pt)--cycle;}},postaction={decorate}},
    ->2/.style={decoration={markings,mark=at position 1 with {\fill (2pt,0)--(-2pt,2.31pt)--(-2pt,-2.31pt)--cycle;}},postaction={decorate}},
}


\def\bra#1{\mathinner{\langle{#1}|}}
\def\ket#1{\mathinner{|{#1}\rangle}}
\newcommand{\Ket}[1]{\vcenter{\hbox{$\displaystyle\stretchleftright{|}{#1}{\bigg\rangle}$}}}
\newcommand*{\Ri}{S}
\newcommand*{\Ric}{S^{\dagger}}
\newcommand{\comment}[1]{{\color{red} (#1)}}

\title{Full Commuting Projector Hamiltonians of Interacting Symmetry-Protected Topological Phases of Fermions}

\author{Nathanan Tantivasadakarn}
\affiliation{Department of Physics, Harvard University, Cambridge, MA 02138, USA}
\author{Ashvin Vishwanath}
\affiliation{Department of Physics, Harvard University, Cambridge, MA 02138, USA}

\date{\today}

\begin{abstract}
Using the decorated domain wall procedure, we construct Finite Depth Local Unitaries (FDLUs)  that realize Fermionic Symmetry-Protected Topological (SPT) phases. This results in explicit `full' commuting projector Hamiltonians, where `full'  implies the fact that the ground state, as well as all excited states of these Hamiltonians, realizes the nontrivial SPT phase.  We begin by constructing explicit examples of 1+1D phases protected by symmetry groups $G=\mathbb Z_2^T \times \mathbb Z_2^F$ , which also has a free fermion realization in class BDI, and $G=\mathbb Z_4 \times \mathbb Z_4^F$, which does not. We then turn to 2+1D, and construct the square roots of the Levin-Gu bosonic SPT phase, protected by $\mathbb Z_2 \times \mathbb Z_2^F$ symmetry, in a concrete model of fermions and spins on the triangular lattice. Edge states and the anomalous symmetry action on them are explicitly derived. Although this phase has a free fermion representation as two copies of $p+ip$ superconductors combined with their $p-ip$ counterparts  with a different symmetry charge, the full set of commuting projectors is only  realized in the strongly interacting version, which also implies that it admits a many-body localized realization. 

\end{abstract}

\maketitle

\tableofcontents

\section{Introduction}\label{intro}
Recently, there has been much progress in classifying ground states of many-body systems with an energy gap. The simplest setting has been the classification of invertible (sometimes also called Short Range Entangled) phases\cite{KitaevKITPTalk, ChenScience, Freed2014}, which feature a unique ground state on a closed manifold. Further, one can consider two cases (i) systems built entirely out of bosons/spins or (ii) those that fundamentally rely on the existence of fermionic excitations. A classification of bosonic Symmetry Protected Topological phases (SPTs), based on the group cohomology of the symmetry group was proposed\cite{Chenetal2013}. While capturing important aspects of the physics, this classification leaves out certain invertible phases of bosons by construction, such as chiral phases in 2+1D which are not protected by any symmetry\cite{Kitaev2006,LuVishwanath2012}. Furthermore, some `beyond cohomology' phases in 3+1D\cite{VishwanathSenthil2013,ChenLuVishwanath2014,Burnelletal2014} are omitted, despite being protected by time reversal symmetry. Various generalized cohomology classifications have been proposed in attempt to capture these missing phases\cite{Kapustin2014,Freed2014,Wen2015,FreedHopkins2016,Xiong2016,GaiottoJohnson-Freyd2017}.

The classification of fermionic invertible phases remains active, despite the fact that free fermion examples of these states, the Chern insulators, have been known since the 1980s \cite{TKNN1982}. However, interactions can play an important role and are known to collapse distinctions between free fermion phases \cite{FidkowskiKitaev2010,GuLevin2014,WangSenthil2014,FidkowskiChenVishwanath2013,Kapustinetal2015} as well as generate entirely new phases that are intrinsically fermionic \cite{WangLinGu2017,ChengTantivasadakarnWang2018}. A pioneering step was taken by Gu and Wen who constructed a supercohomology classification\cite{GuWen2014}, in close analogy with the cohomology construction, although several phases and symmetry groups had to be excluded in this formulation. For example, only symmetry groups of the form $G\times \mathbb Z_2^F$ were permitted, where the $\mathbb Z_2^F$ factor corresponds to fermion parity. However, some of the most interesting cases correspond to symmetry groups that cannot be represented in this way. For example, in electronic systems with Kramers degeneracy, time reversal squares to the fermion parity  $T^2=P_f$, implying that one cannot decompose the symmetry as a simple product. Subsequently, spin TQFTs and other generalized cohomology classifications have emerged\cite{Freed2014,Chengetal2015,Kapustinetal2015,FreedHopkins2016,Xiong2016,KapustinThorngren2017,GaiottoJohnson-Freyd2017}.

The classification of Symmetry-Protected Topological (SPT) phases has gone hand in hand with the construction of exactly soluble models for both bosons\cite{AKLT1987,ChenLiuWen2011,LevinGu2012,Chenetal2013,ChenLuVishwanath2014,Santos2015,Yoshida2015,Yoshida2016,ZengZhou2016,GeraedtsMotrunich2014,Tsuietal2017,ChenPrakashWei2018} and fermions\cite{GuWen2014,TarantinoFidkowski2016,Bhardwajetal2017,WangNingChen2017,WangGu2018}. It is worth noting the distinct ways in which a model can be exactly soluble. In some cases, one can solve exactly for the ground state of a Hamiltonian, but not for the general excited states. This is the situation for the Affleck-Kennedy-Lieb-Tasaki \cite{AKLT1987} model which realized the 1D Haldane phase of the spin-one chain, one of the first SPTs to be discovered\cite{Haldane1983}. In other cases, one can obtain the ground state and those excited states that lie within a certain constrained subspace, but not all states in the Hilbert space. Here, we will be interested in exactly soluble Hamiltonians that capture the ground state as well as excited states that span the entire Hilbert space, which we will term Full Commuting Projector Hamiltonians (FCPHs).

There are several reasons to focus on FCPHs. Obviously, they give us analytical control over deriving universal properties of topological phases. Moreover, for fermionic SPTs that also possess a free fermion description, this alternate starting point can help access nonperturbative effects that are obscured in the free fermion limit. Indeed we will discuss a 2+1D fermion SPT, for which interactions will enable us to write a commuting projector Hamiltonian, which is not possible in the free fermion limit.  

Furthermore, they allow us to extend the discussion of topology to states beyond the ground state. Since every excited energy-eigenstate is also an eigenstate of the commuting projectors, they can all be said to be in the same SPT phase, and the topological properties are now associated with the entire Hamiltonian rather than just with its ground state. In the presence of disorder and Many-Body Localization (MBL), one can potentially extend these properties from specific Hamiltonians to actual phases that are stable to the addition of any local symmetry preserving perturbation. The resulting MBL Hamiltonians then feature SPT order \cite{Huseetal2013,Bahrietal2015,Chandranetal2014,PotterVishwanath2015}.  On the practical side, this may help in the realization of SPT phases in quantum systems where cooling to the ground state may be challenging, but in the presence of MBL, no cooling would be required. Rather, the SPT physics would reveal itself for example in the coherent quantum dynamics of the edge modes \cite{Bahrietal2015}. 

While MBL is often discussed in terms of the stability of the fully localized free fermion Anderson insulator to the addition of interactions, here we will discuss a fermionic SPT that we believe can be Many-Body Localized in the presence of strong interactions, but has no fully localized free fermion analog. This represents another avenue for a qualitatively new effects enabled by interactions. Specifically, we will discuss a 2+1D fermionic SPT phase consisting of two copies of $p+ip$ superconductors, combined with two copies of $p-ip$ superconductors, which transform in opposite ways under a global $\mathbb Z_2$ symmetry. At the free fermion level, one cannot write down a set of localized Wannier orbitals (the free fermion analog of commuting projectors) that also preserve symmetry for such a topological band. Nevertheless, we will see that this phase can be represented with a FCPH with interactions. 

Looking forward, we expect that identifying physical obstructions to realizing commuting projector models can provide insight into the classification of topological phases and intrinsic differences. For example, it has been argued that invertible phases with edge modes displaying a net chirality in 2+1D cannot be captured by a commuting projector Hamiltonian \cite{Kitaev2006,LinLevin2014,PotterVishwanath2015}. It would be interesting to know which non-chiral phases can be realized by FCPHs and which ones are fundamentally obstructed.   

Let us briefly review related earlier work. In Ref. \onlinecite{GuWen2014}, Gu \& Wen constructed lattice Hamiltonians for a certain class of fermionic SPTs with groups of the form $G \times \mathbb Z_2^F$. However, the FDLU they constructed is only unitary if the creation and annihilation operators are changed to Majorana operators. In this paper, we are able to construct a FCPH for a 2D SPT with $\mathbb Z_2 \times \mathbb Z_2^F$ symmetry, where the properties of the phase, such as the ground state wavefunction, has an intuitive form. Furthermore, we also construct other fermionic SPTs whose symmetry group is not of the form $G \times \mathbb Z_2^F$, and we outline a procedure that could possibly be used to construct SPTs for a large class of symmetries in up to three dimensions.

We present three concrete examples. The first is a 1D fermionic SPT protected by time reversal symmetry (class BDI), where we introduce the Fermionic version of the Decorated Domain Wall procedure\cite{ChenLuVishwanath2014}(although full commuting projector models exist from the free theory). Our second example is a 1D fermionic SPT protected by $\mathbb Z_4 \times \mathbb Z_4^F$, the generating phase of which is intrinsically fermionic and intrinsically interacting. Furthermore this symmetry group, which does not split into $G\times \mathbb Z_2^F$, is outside of supercohomology. Finally, we present a model of the Fermionic Ising SPT in 2D ($\mathbb Z_2 \times \mathbb Z_2^F$), which is the main result of this paper. The classification of these phases are summarized in Table \ref{tab:classification} In all three examples, we discuss the ground state wave function, we show that the FDLU, if applied twice, gives us the bosonic embedded SPT, and we show the non-trivial properties of the edge states.

Our paper is organized as follows. In Section \ref{prelim}, we review the concepts of SPTs, commuting projectors, and local unitary gates used to construct these models. We then present our three models in Sections \ref{1D}, \ref{intrinsic}, and \ref{2D}, respectively. We argue how to construct phases with more general symmetries in Section \ref{general}. Finally, in Section \ref{MBL}, we prove that all the eigenstates of the FCPH are in the same phase,  and remark on the viability of MBL. 

While this paper was in preparation, we recently learnt of a work by Ellison \& Fidkowski\cite{EllisonFidkowski2018}, who constructed FDLUs for all supercohomology phases in 2D with explicit spin structure dependence in the construction. In particular, their unitary can be used to obtain an alternative FCPH for the $\mathbb Z_2 \times \mathbb Z_2^F$ SPT in 2D. It would be interesting to see how our two models are related.

\begin{table}
\caption{Classification of SPT phases in consideration}
\begin{tabular}{|c|c|c|c|}
\hline
SPT &$\mathbb Z_2^T \times \mathbb Z_2^F$ 1D  & $\mathbb Z_4 \times \mathbb Z_4^F$ 1D & $\mathbb Z_2 \times \mathbb Z_2^F$ 2D\\
\hline
Classification & $\mathbb Z_8$ & $\mathbb Z_4$ &  $\mathbb Z_8$\\
Cohomology & $\nu=0,4$ & $\nu=0,2$ & $\nu=0,4$\\
Supercohomology &$\nu=0,2,4,6$ & N/A & $\nu=0,2,4,6$\\
Models we consider &$\nu=2,6$ & $\nu=1,3$ & $\nu=2,6$\\
\hline
\end{tabular}
\label{tab:classification}
\end{table}

\section{Preliminaries}
\label{prelim}
\subsection{SPT Phases and Local Unitaries}
We review the nominal definitions of SPT phases that we will be using in the paper. For a more detailed and rigorous overview, see for example Refs. \onlinecite{ChenGuWen2010,GuWangWen2015,HuangChen2015}. Here, we always assume that our spatial dimension is greater than zero.  We define a Finite-Depth Local Unitary (FDLU) to be a quantum circuit
\begin{equation}
U = \prod_{i_M} U_{i_M}^{(M)} \prod_{i_{M-1}} U_{i_{M-1}}^{(M-1)} \cdots \prod_{i_1} U_{i_1}^{(1)}
\end{equation}
consisting of a finite number of layers $M$, where each layer is made up of a product of unitary gates $U_{i_{m}}$ that are local and all commute within each layer.

Now, let our system have some symmetry $G$, represented by some faithful unitary representation which acts onsite $R(g)$. We define a gate to be symmetric if it commutes with $R(g)$. We now say that the FDLU is \textit{symmetric} if all the gates commute with our symmetry. That is,
\begin{equation}
[R(g), U_{i_m}] = 0.
\end{equation}
Two gapped Hamiltonians related by a symmetric FDLU are in the same phase. On the other hand, we call the FDLU \textit{overall symmetric} if the whole FDLU still commutes with $R(g)$,
\begin{equation}
[R(g), U ] =0.
\end{equation}
Actually, we can relax this definition and allow for them to commute up to one-dimensional representation operators. i.e. 0D SPTs. As an example, for $G=\mathbb Z_2$, a gate is symmetric if it commutes or anticommutes with $R(g)$.

Consider all gapped Hamiltonians which have a unique ground state on a closed manifold. We can now define SPT phases as an equivalence class of these Hamiltonians under symmetric FDLUs. For a non-trivial SPT Hamiltonian, we therefore see that there does not exist a symmetric FDLU that connects it to a product state Hamiltonian, and so the FDLU can be at most overall symmetric.

For fermionic SPTs, we have an additional constraint. Since fermion parity $P_f$ is never explicitly broken, all gates must commute with fermion parity. In other words, the FDLU must be symmetric with respect to $P_f$ but need only be overall symmetric with respect to the remaining symmetries.

\subsection{Symmetric Local FCPHs}
To construct SPT phases using FCPHs, the projectors must also be local and symmetric. Thus, we define a symmetric local FCPH as a Hamiltonian
\begin{align}
\label{equ:dressedHformal}
H &= - \sum_{i,\alpha} h_i^\alpha \Pi_i^\alpha,\\
\Pi_i^\alpha &= UP_i^\alpha U^\dagger,
\end{align}
where $h_i^\alpha$ are real, $i$ denotes an index running over the local Hilbert spaces, and $\alpha$ denotes the irreducible representations of the symmetry. We demand the following properties:\\
1. $U$ is an overall symmetric FDLU.\\
2. $P_i^\alpha$ are projectors ($(P_i^\alpha)^2 = P_i^\alpha$) with unit rank in the onsite Hilbert space.\\
3. $[P_i^\alpha,P_j^\beta]=0$ for all $i,j,\alpha,\beta$.\\
4. Let $R(g)_i^\alpha$ be the representation of the onsite symmetry at site $i$; then $[R(g)_i^\alpha,P_i^\alpha]=0$.\\
5. $\sum_{\alpha} P_i^\alpha=\mathbbm 1_i$ where $\mathbbm 1_i$ is the identity operator in the onsite Hilbert space.

Henceforth, we will call $P_i^\alpha$ and $\Pi_i^\alpha$ the \textit{bare projectors} and \textit{dressed projectors} respectively. We can also define the \textit{bare Hamiltonian}
\begin{equation}
\label{equ:bareHformal}
H_0 = - \sum_{i,\alpha} h_i^\alpha P_i^\alpha,
\end{equation}
so that the \textit{dressed Hamiltonian} Eq. \eqref{equ:dressedHformal} can be written as a unitary evolution
\begin{equation}
H = UH_0U^\dagger.
\end{equation}
 The role of the FDLU $U$ is to disentangle the dressed projectors into bare projectors, which are onsite. The Hamiltonian defined is a sum of dressed projectors that form a full set of symmetric commuting local observables for the entire Hilbert space.

We now restrict our attention to the symmetry in discussion. Our Hilbert space is defined on a lattice, with local degrees of freedom being qubits living on the vertices $i \in V$, and fermions living on the vertices of the dual lattice $l \in V^*$ (we defer the discussion of qudits and $\mathbb Z_4^F$ representations to Section \ref{intrinsic}). We choose our $\mathbb Z_2$ symmetry to be represented by $\chi = \prod_i X_i$ for $\mathbb Z_2$ or $T = \prod_i X_i K$ for $\mathbb Z_2^T$, where $K$ is complex conjugation.
For the qubits, the onsite symmetry $X_i$ has two irreducible representations: the $\ket{+}_i$ and $\ket{-}_i$ states transform trivially and non-trivially under the symmetry, respectively. The corresponding projectors are 
\begin{align}
P_i^+ &= \frac{1+X_i}{2}, & P_i^- &= \frac{1-X_i}{2}.
\end{align}
However, suppose we only wish to discuss the ground state of the bare Hamiltonian \eqref{equ:bareHformal}, then for each local site, we can always shift the Hamiltonian by some constant so that $h_i^-=0$. For this reason, we only need to write down $P_i^+$ in our Hamiltonian. Note that in the literature, the ``projectors" are often defined as $X_i$, despite actually being involutory matrices (squaring to identity instead of to itself). Though this simplifies the form of the Hamiltonian, we stick to the original definition for clarity.

For fermions, our fermion parity operator acts as
\begin{equation}
P_f =\prod_{l} (-1)^{n_{l}}
\end{equation}
where $n_{l} = c_{l}^\dagger c_{l}$ is the number operator of the fermion at the dual vertex $l$.
There are two projectors, corresponding to the occupied and unoccupied states, but similarly, we can always shift our Hamiltonian so that we only write down the unoccupied state
\begin{equation}
P_{l}^0 = 1-n_{l} = \frac{1}{2}(1+(-1)^{n_{l}}).
\end{equation}

To summarize, our bare Hamiltonian has the form
\begin{equation}
H_0 = - \sum_{i \in V} h_i^+ \frac{1}{2}(1+X_i) - \sum_{l \in V^*} h_{l}^0 \frac{1}{2}(1+(-1)^{n_{l}}).
\label{equ:bareH}
\end{equation}
 The evolved Hamiltonian can thus be written in terms of dressed involutories
\begin{equation}
H = - \sum_{i \in V} h_i^+ \frac{1}{2}(1+\bar X_i) - \sum_{l \in V^*} h_{l}^0 \frac{1}{2}(1+(-1)^{\bar{n}_{l}}).
\label{equ:dressedH}
\end{equation}
where
\begin{align}
\bar X_i &= U X_i U^\dagger & (-1)^{\bar{n}_{l}} &= U (-1)^{n_{l}} U^\dagger.
\end{align}

In the next few sections, we set all coefficients $h_i^+$ and $h_{l}^0$ to one for simplicity, as it does not change the ground state. We will later restore the full set of commuting projectors in Section \ref{MBL} to discuss the excited states.
\begin{table*}
\caption{Definitions of quantum gates for qubits and fermions}
\begin{tabularx}{\textwidth}{|c|r l|X|}
\hline
Name & \multicolumn{2}{c|}{Action}  &Remark\\
\hline
Pauli $X$ & $X_i \ket{g_i}$ &$= \ket{1-g_i}$&\\
Pauli $Z$ & $Z_i \ket{g_i}$ &$= (-1)^{g_i}\ket{g_i}$&\\
$\frac{\pi}{2}$ phase gate ($S$) & $\Ri_i \ket{g_i}$ &$= i^{g_i}\ket{g_i}$& $S_i^2=Z_i$.\\
Controlled-$Z$ & $CZ_{ij} \ket{g_i,g_j}$ &$= (-1)^{g_ig_j}\ket{g_i,g_j}$ & $CZ_{ij}=CZ_{ji}$.\\
Controlled-$\Ri$ & $C\Ri_{ij} \ket{g_i,g_j}$ &$= i^{g_ig_j}\ket{g_i,g_j}$ & $CS_{ij}=CS_{ji}$,$(CS_{ij})^2=CZ_{ij}$. \\
Controlled-controlled-$Z$ & $CCZ_{ijk} \ket{g_i,g_j,g_k}$ &$= (-1)^{g_ig_jg_k}\ket{g_i,g_j,g_k}$&\\
Controlled-controlled-$\Ri$ & $CC\Ri_{ijk} \ket{g_i,g_j,g_k}$ &$= i^{g_ig_jg_k}\ket{g_i,g_j,g_k}$&\\
\hline
Majorana & $\gamma_{l} \ket{f}$ &$= (c_{l} + c_{l}^\dagger)\ket{f}$ & \\
		& $\tilde \gamma_{l}\ket{f}$ &$= i(c_{l} - c_{l}^\dagger)\ket{f}$ & Fermion parity can be written as $n_l = i \gamma_l \tilde \gamma_l$\\
Controlled-Majorana	& $C_i \gamma_{l}  \left (\ket{g_i} \otimes \ket{f} \right)$ & $= \ket{g_i} \otimes \gamma_{l}^{g_i} \ket{f}$& Analog of CNOT where controlled site is a fermion instead of a qubit.\\
Controlled-controlled-Majorana & $CC_{ij} \gamma_{l} \left (\ket{g_i,g_j} \otimes \ket{f} \right)$ &$= \ket{g_i,g_j} \otimes \gamma_{l}^{g_ig_j} \ket{f}$&\\
\hline
\end{tabularx}
\label{tab:gates}
\end{table*}
\subsection{Quantum Gates}
In this paper, our FDLUs are constructed using quantum gates. We define these gates in Table \ref{tab:gates} according to their action on the qubits and fermions. Here, $g_i=0,1$ labels states in the computational basis and $\ket{f}$ labels some state in the fermionic Fock space.

Throughout the paper, we will sometimes abuse notation and use $\gamma_{l}^{g_i}$ to represent $C_i \gamma_{l}$. As an example, this cleans up expressions and allows us to write $\gamma_{l}^{g_i g_j+g_k}$ instead of $CC_{ij}\gamma_{l} C_k\gamma_{l}$. Similarly, $CCZ_{ijk}$ and $CC \Ri_{ijk}$ may sometimes be interchanged with $(-1)^{g_ig_jg_k}$ and $i^{g_ig_jg_k}$ respectively. Nevertheless, one must be careful with expressions like $\gamma_{l}^{g_i} X_i$, where the $g_i$ in the exponent cannot be taken literally when written in front of an $X_i$ operator. For this example, we instead have 
\begin{align}
\gamma_l^{g_i} X_i \left (\ket{g_i} \otimes \ket{f}\right ) &= C_i \gamma_l\ket{1-g_i} \otimes  \ket{f} \nonumber\\
&=\ket{1-g_i} \otimes \gamma_l^{1-g_i} \ket{f}.
\end{align}

\section{Model for 1D SPT with $G=\mathbb Z_2^T \times \mathbb Z_2^F$}
\label{1D}
In this section, we present the fermionic decorated domain wall unitary and use it to construct commuting projectors for topological superconductors in 1D. Although commuting projector models are known to exist even in the free case (i.e. Majorana chains), we present a construction of the interacting case which is generalizable to arbitrary dimensions.

The model we will present is a 1D Fermionic SPT protected by $G=\mathbb Z_2^T \times \mathbb Z_2^F$. It is known that free topological superconductors (class BDI) are classified by $\mathbb Z$ in 1D with topological index $\nu$ generated by stacking Majorana chains. This classification collapses to $\mathbb Z_8$ under interactions\cite{FidkowskiKitaev2011}. Furthermore, four copies ($\nu=4$) can be obtained by embedding the cluster state\cite{Sonetal2012,Santos2015,Yoshida2015,Yoshida2016,ZengZhou2016} protected by time reversal symmetry into the trivial phase $\nu=0$. In this discussion, we focus only on the supercohomology phases, which have even index $\nu$.

\subsection{Ground State Wave Function for $\nu=2$}
We begin by discussing the unitary, which in turn determines the ground state wave function. Let us first recall the procedure in the bosonic case using the cluster state. The Hamiltonian that realizes this as its ground state is
\begin{equation}
H = - \sum_{i=1}^{2N} \frac{1}{2}\left(1+Z_{i-1}X_i Z_{i+1} \right),
\end{equation}
which we will call the cluster Hamiltonian. This Hamiltonian and the cluster state can be obtained by evolving respectively the bare projector and the product state of all $\ket{\rightarrow}$ with the unitary
\begin{equation}
U = \prod_{i=1}^{2N} CZ_{i,i+1}.
\end{equation}
If we define two $\mathbb Z_2$ symmetries as the product of $X$ operators on the odd and even sites, respectively, we see that the unitary creates the cluster state by attaching the charge of the second symmetry on the domain walls of the first symmetry\cite{ChenLuVishwanath2014} as depicted in Figure \ref{fig:DDW1Dboson}.  A more thorough review of the procedure is treated in Appendix \ref{cluster}. We will now import this procedure to construct fermionic SPTs.

\begin{figure}
\includegraphics{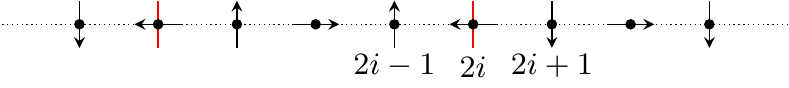}
\caption{Decorated Domain Wall for bosons. A non-trivial $\mathbb Z_2$ SPT $\ket{\leftarrow}$ at site $2i$ is created if there is a domain wall (red line) between the qubits at sites $2i-1$ and $2i+1$}
\label{fig:DDW1Dboson}
\end{figure}

Our system is defined on a ring of $N$ sites, where qubits are labeled $i$ and fermion sites live on vertices in the dual lattice, which we label with an index $i+\frac{1}{2}$. Our bare Hamiltonian Eq. \eqref{equ:bareH} can then be written as
\begin{equation}
H_0=-\sum_{i=1}^N \left[\frac{1}{2}(1+X_i) +\frac{1}{2}(1+(-1)^{n_{i+\frac{1}{2}}}) \right ].
\end{equation}

\begin{figure}
\includegraphics{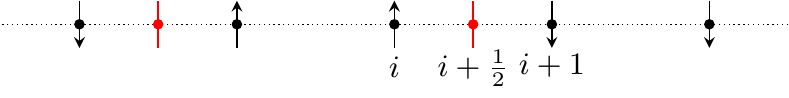}
\caption{Decorated Domain Wall for fermions. The fermion at site $i+\frac{1}{2}$ is excited (red dot) if there is a domain wall (red line) between the qubits at sites $i$ and $i+1$. The creation operators are written in order from right to left to obtain the $\nu=2$ phase, and in the reversed order for the $\nu=6$ phase.}
\label{fig:DDW1D}
\end{figure}

Recall that Fermionic invertible phases in 0D are classified by $\mathbb Z_2$, corresponding to even and odd parity (or empty and occupied states respectively). The two states can be connected by the Majorana operator $\gamma$, and so our unitary can create a non-trivial $\mathbb Z_2^T \times \mathbb Z_2^F$ SPT by applying the controlled-Majorana to enforce the decoration of fermions on the $\mathbb Z_2^T$ domain walls as shown in Figure \ref{fig:DDW1D}. Note that since there is always an even number of domain walls on a ring, the evolution operator commutes with fermion parity. However, unlike the bosonic case, the local gates do not commute. In this case, we must choose an ordering for our decoration procedure. Here, we choose to excite the fermions from left to right (that is, the operators are written from right to left). In conclusion, our unitary can be written similarly to the cluster state unitary
\begin{align}
 U&=Z_1 \prod_{i=N}^1   C_i\gamma_{i+\frac{1}{2}} C_{i+1}\gamma_{[i+1]-\frac{1}{2}} \nonumber\\
 &=Z_1 ( \gamma_{N+\frac{1}{2}}^{g_N} \gamma_{\frac{1}{2}}^{g_1}) \gamma_{N-\frac{1}{2}}^{g_{N-1}+g_N} \cdots \gamma_{1+\frac{1}{2}}^{g_1+g_2}.
\label{equ:U_1D}
 \end{align}
Here, we assume periodic boundary conditions for the qubits ($g_{N+1} = g_1$) and $[i+1]$ is defined modulo $N$. There are two reasons why we must include the extra $Z_1$ in the unitary. First, it is needed so that the form of $U$ is invariant under shifting all the indices by one site. Indeed, moving $\gamma_{1+\frac{1}{2}}^{g_1+g_2}$ to the front gives a factor of $(-1)^{g_1+g_2}=Z_1Z_2$ and shifting the indices $i \rightarrow i-1$ recovers the same expression for $U$. Second, we need to determine the boundary conditions for the fermions, which is also why the first term in the product is written as two different Majoranas. This boundary condition dependence of the unitary reflects the spin structure dependence of Fermionic SPTs\cite{GaiottoKapustin2016}. There are two boundary conditions we can choose.\\
1. Antiperiodic boundary conditions $\gamma_{\frac{1}{2}}=-\gamma_{N+\frac{1}{2}}$ which corresponds to the trivial or Neveu-Schwarz (NS) spin structure. In this case, we have
\begin{align}
 U_{NS}=\gamma_{N+\frac{1}{2}}^{g_N+g_1} \gamma_{N-\frac{1}{2}}^{g_{N-1}+g_N} \cdots \gamma_{1+\frac{1}{2}}^{g_1+g_2}.
\label{equ:U_NS}
 \end{align}
 2. Periodic boundary conditions: $\gamma_{\frac{1}{2}}=\gamma_{N+\frac{1}{2}}$, which corresponds to the non-trivial or Ramond (R) spin structure, in which case
 \begin{align}
 U_{R}=Z_1 \gamma_{N+\frac{1}{2}}^{g_N+g_1} \gamma_{N-\frac{1}{2}}^{g_{N-1}+g_N} \cdots \gamma_{1+\frac{1}{2}}^{g_1+g_2}.
\label{equ:U_NS}
 \end{align}

One might notice that since the gates do not commute, the unitary defined above actually has linear depth, and therefore is not an FDLU. To remedy this problem, we enlarge our Hilbert space by introducing an additional fermion site at each position $i$ with Majorana operators $\eta_i$ and $\tilde \eta_i$ respectively. We claim that by doing so, we can rewrite our unitary in Eq. \eqref{equ:U_1D} as
\begin{align}
 U =&  \prod_{i=N}^1 \left [ \tilde \eta_{i}^{g_i} \eta_{i}^{g_i}  \right]   \prod_{i=N}^1 \left [ \tilde \eta_{i+1}^{g_{i+1}}  \gamma_{i+\frac{1}{2}}^{g_{i}} \gamma_{[i+1]-\frac{1}{2}}^{g_{i+1}} \eta_{i}^{g_i}\right],
 \label{equ:U_1Dfinitedepth}
\end{align}
where we have grouped terms into gates written in square brackets. Each gate always changes the number of fermions by an even number. Furthermore, all the gates are local, and commute with one another within each product. Thus, $U$ written as shown is a symmetric fermionic FDLU. If a spin structure is chosen, the expression simplifies to
\begin{align}
 U_{NS} =& &Z_1 &\prod_{i=N}^1 \left [ \tilde \eta_{i}^{g_i} \eta_{i}^{g_i}  \right]   \prod_{i=N}^1 \left [ \tilde \eta_{i+1}^{g_{i+1}} \gamma_{i+\frac{1}{2}}^{g_{i}+g_{i+1}} \eta_{i}^{g_i}\right],\\
  U_{R} =&  & &\prod_{i=N}^1 \left [ \tilde \eta_{i}^{g_i} \eta_{i}^{g_i}  \right]   \prod_{i=N}^1 \left [ \tilde \eta_{i+1}^{g_{i+1}} \gamma_{i+\frac{1}{2}}^{g_{i}+g_{i+1}} \eta_{i}^{g_i}\right].
\end{align}
Notice that the boundary conditions of $\eta$ and $\tilde \eta$ do not affect the expression of $U_{NS}$ or $U_{R}$. We remark that the supercohomology construction with choice of cocycles $\omega(a,b)= (-1)^{ab}$ and $\beta(a)=a \text{ (mod 2)}$ gives $U_R$, but cannot reproduce $U_{NS}$.

Now, we show that the unitary given in Eq. \eqref{equ:U_1Dfinitedepth} is equivalent to the previous one. First, we move the first term in the second product  ($\tilde \eta_{1}^{g_1}$) to the back. This gives a phase factor of $(-1)^{g_1}$, which is $Z_1$ in Eq. \eqref{equ:U_1D}. Doing so, we see that the second product can be rewritten as
\begin{equation}
\prod_{i=N}^1  \gamma_{i+\frac{1}{2}}^{g_{i}} \gamma_{[i+1]-\frac{1}{2}}^{g_{i+1}} \eta_{i}^{g_{i}}\tilde \eta_{i}^{g_{i}}.
\end{equation}
The terms $\eta_{i}^{g_{i}}\tilde \eta_{i}^{g_{i}}$ commute with all other terms in the second product, so we can move them to the front and cancel all the terms in the first product, and we are left with the original evolution operator.

\subsection{Commuting Projector for $\nu=2$}
\label{1Dmodel}
With the unitary, we can now obtain the commuting projector. First, conjugating the $\mathbb Z_2^T$ symmetry with the unitary, we find
\begin{equation}
U T U^\dagger = -T \gamma_{\frac{1}{2}}\gamma_{N+\frac{1}{2}}.
\end{equation}

Now we evolve the Hamiltonian. We find that the dressed $X$ operators are
\begin{equation}
\bar X_i = Z_i\gamma_{i-\frac{1}{2}}X_i\gamma_{i+\frac{1}{2}}Z_{i+1}.
\end{equation}
Thus we conclude that for antiperiodic boundary conditions then $UTU^\dagger =T$, but our Hamiltonian is not uniform. On the other hand, if we choose periodic boundary conditions, then $UTU^\dagger =-T$, but our Hamiltonian is uniform. Here, we can see that the antiperiodic boundary conditions is considered ``trivial" in the sense that the ground state wave function transforms under the trivial $\mathbb Z_2^T$ representation, while the non-trivial spin structure gives a wave function that transforms under the non-trivial $\mathbb Z_2^T$ representation.

To dress the fermions, we use the fact that $\gamma_{i+\frac{1}{2}}$ is only applied on the domain wall where $Z_iZ_{i+1}=-1$, in which case
\begin{equation}
 \gamma_{i+\frac{1}{2}} n_{i+\frac{1}{2}} \gamma_{i+\frac{1}{2}} =1-n_{i+\frac{1}{2}}.
\end{equation}
Doing so, we find that the dressed fermion parity operator is
\begin{align}
(-1)^{\bar n_{i+\frac{1}{2}}}&= Z_i  (-1)^{n_{i+\frac{1}{2}}}Z_{i+1}.
\end{align}

To summarize, our dressed Hamiltonian for the R spin structure is
\begin{align}
H_2=-\sum_i &\left[ \frac{1}{2}(1+\gamma_{i-\frac{1}{2}} Z_i X_i  \gamma_{i+\frac{1}{2}}Z_{i+1} )\right. \nonumber\\
 & \ \left.+\frac{1}{2}(1+Z_i (-1)^{n_{i+\frac{1}{2}}} Z_{i+1}) \right],
 \label{equ:H21D}
\end{align}
while for the NS spin structure, the term with $X_1$ has a negative coefficient.

The dressed operators in the above Hamiltonian have the following physical interpretation. $(-1)^{\bar n_{i+\frac{1}{2}}} =Z_i (-1)^{n_{i+\frac{1}{2}}} Z_{i+1}$ binds the excited fermions ($\mathbb Z_2^F$ charges) to the domain walls, while $\bar X_i = \gamma_{i-\frac{1}{2}} Z_i X_i  \gamma_{i+\frac{1}{2}}Z_{i+1}$ creates or annihilates domain wall--fermion bound pairs and fluctuates them.

\subsection{$\nu=2$ is the ``Square Root" of the Bosonic Phase}
Next, we will show that evolving with $U^2$ gives the negative of the cluster Hamiltonian with bare fermion projectors. First, we notice that $U^2$ has no fermionic operators
\begin{align}
U^2&= Z_1^2\left ((\gamma_{N+\frac{1}{2}}^{g_N}\gamma_{\frac{1}{2}}^{g_1} )  \gamma_{N-\frac{1}{2}}^{g_{N-1}+g_N} \cdots \gamma_{1+\frac{1}{2}}^{g_1+g_2}\right )^2 \nonumber\\
&=(-1)^{\sum_{j<k} (g_j+g_{j+1}) (g_k+g_{k+1})}.
\label{equ:1DsignU2}
\end{align}
As a result, $U^2$ will leave the fermions undressed. We also notice that $U^2$ does not depend on the boundary conditions of the fermions. From the expression, we see that $U^2$ is simply a product of $CZ$ operators, so we can compute $U^2X_iU^{\dagger2}$. This is done explicitly in Appendix \ref{1Dapp}, and we find
\begin{equation}
U^2X_iU^{\dagger2} = -Z_{i-1}X_iZ_{i+1}.
\end{equation}
Hence, the Hamiltonian is
\begin{align}
H_4&= U^2H_0U^{\dagger^2} \nonumber\\
&=-\sum_i \left[ \frac{1}{2}(1- Z_{i-1} X_i  Z_{i+1} ) +\frac{1}{2}\left(1+(-1)^{n_{i+\frac{1}{2}}}\right) \right],
\end{align}
which is in the $\nu=4$ phase. This is just the negative of the cluster Hamiltonian with bare fermions, and so is bosonic in the sense that it does not depend on the spin structure. The ground state wave function gives a minus sign to each spin down region. This can also be seen from the action of $U^2$ on the product state. Since $U$ attaches fermions to the domain walls (the boundary of each spin down region), applying $U$ again gives $-1$ for each spin down region from anticommuting the pair of Majorana operators to square them away.

 Similarly, evolving with $U^3$ gives
\begin{align}
H_6= U^3H_0U^{\dagger3}=-\sum_i &\left[ \frac{1}{2}(1-Z_{i-1}\gamma_{i-\frac{1}{2}} Z_i X_i  \gamma_{i+\frac{1}{2}} ) \right. \nonumber\\
 & \left.+\frac{1}{2} \left (1+Z_i (-1)^{n_{i+\frac{1}{2}}} Z_{i+1}\right) \right  ]
 \label{equ:nu6_1D}
\end{align}
for the R spin structure.

\subsection{Projective Representation of the Edge}
To further confirm that our models are indeed non-trivial SPT phases, we can check that the edges of an open chain transform under projective representations of the symmetries. These are given for phases with even $\nu$ in Table \ref{tab:projrep1D}. For a derivation of these properties, see Refs. \onlinecite{FidkowskiKitaev2011,VerresenMoessnerPollmann2017}. We remind the reader that for 1D fermionic SPTs, the projective properties on the left and right edges need not be the same. In fact, the properties on the right edge for a phase with index $\nu$ are the same as the properties on the left edge for a phase with index $-\nu$. 

\begin{table}
\caption{Projective representation of the symmetries on the left edge of the $\mathbb Z_2^T \times \mathbb Z_2^F$ SPT in 1D}
\begin{tabular}{|c|c|c|c|}
\hline
$\nu$ & $T^2$   & $(TP_f)^2$\\
\hline
0 & 1   & 1\\
2 & 1   & -1\\
4 & -1   & -1\\
6 & -1   & 1\\
\hline
\end{tabular}
\label{tab:projrep1D}
\end{table}

Let us compute this explicitly for $\nu=2$. The dressed projectors in Eq. \eqref{equ:H21D} must be equal to one when acting on the ground state. From this, we see that the action of the local symmetry operators satisfy
\begin{align}
X_i &= - Z_i Z_{i+1} \gamma_{i-\frac{1}{2}}\gamma_{i+\frac{1}{2}}, \\
(-1)^{n_{i+\frac{1}{2}}}  &= Z_i Z_{i+1}.
\end{align}
Inserting these into the expression for the symmetries, we obtain
\begin{align}
T &= X_1 Z_2  \gamma_{3/2} \gamma_{N-\frac{1}{2}} Z_{N} X_N K, \\
P_f&= Z_1 Z_N (-1)^{n_{N+\frac{1}{2}}},\\
T P_f &= X_1 Z_1 Z_2 \gamma_\frac{3}{2} \gamma_{N-\frac{1}{2}} X_N (-1)^{n_{N+\frac{1}{2}}}.
\end{align}
Restricting the symmetries to the left side, we find the projective representations $T^2=1$, $P_f^2=1$, $(TP_f)^2=-1$ in agreement with Table \ref{tab:projrep1D}.

Similarly, one can verify the projective representations at the edge for the $\nu=6$ Hamiltonian in Eq.\eqref{equ:nu6_1D}. However, an easier way to see this is by observing that we recover $H_2$ upon applying a symmetric FDLU $\prod_i Z_i$, which negates all the qubit terms, and reflecting the $\nu=6$ Hamiltonian in Eq. \eqref{equ:nu6_1D} from left to right. Thus we see that must be in the $\nu=-2=6$ phase.

We would like remark that our unitary and Hamiltonian can also be viewed as a non-trivial SPT protected by $\mathbb Z_2 \times \mathbb Z_2^F$ where we make the $\mathbb Z_2^T$ symmetry unitary by removing complex conjugation. One can verify that the projective representations for the $\nu=2$ Hamiltonian do correspond to the $\mathbb Z_2 \times \mathbb Z_2^F$ supercohomology phase. Alternatively, one can also show this by computing a topological invariant, which is shown in Appendix \ref{topinv}. In the next section, we will generalize this model to the case where fermion parity forms a semi-direct product with one of the group elements.

\section{Model for 1D SPT with $G=\mathbb Z_4 \times \mathbb Z_4^F$}
\label{intrinsic}
For our first generalization, we construct the generating SPT phase with $G=\mathbb Z_4 \times \mathbb Z_4^F$ symmetry, which has a $\mathbb Z_4$ classification. Here, the group $\mathbb Z_4^F=\mathbb Z_2^F \rtimes \mathbb Z_2$ is generated by a group element $\chi$ which squares to fermion parity. Supercohomology does not provide a classification for such symmetry group, and the generating phase is known to be intrinsically fermionic and intrinsically interacting\cite{FidkowskiKitaev2011,WangLinGu2017}. 

Since our unitary will decorate a 0D $\mathbb Z_4^F$ SPT on $\mathbb Z_4$ domain wall, we first need to discuss how to construct $\mathbb Z_4^F$ SPT phases, which are classified by $\mathbb Z_4$.

\subsection{$\mathbb Z_4^F$ SPT in 0D}
We have a four-dimensional Hilbert space made of a qubit and a fermion. We can label the Hilbert space by $\ket{0}, \ket{1}, c^\dagger\ket{0}, c^\dagger\ket{1}$. As usual, we can define Majorana operators $\gamma = c+ c^\dagger$ and $\tilde \gamma =i(c-c^\dagger)$, and the fermion parity operator $(-1)^n = i \gamma \tilde\gamma$ for the fermion.
Let us define our $\mathbb Z_2$ symmetry $\chi$ as
\begin{equation}
\chi =  X S C\gamma C\tilde \gamma =X (-1)^{ng}.
\end{equation}
where $g=0,1$. We see that
\begin{align}
\chi^2 &=  X S X S \gamma^{1-g} \tilde \gamma^{1-g} \gamma^{g} \tilde \gamma^{g} \nonumber= i \gamma \tilde \gamma = (-1)^n = P_f,
\end{align}
so the $\mathbb Z_2$ symmetry squares to fermion parity as desired.
The eigenvalues of $\chi$ are $ \pm 1, \pm i$ corresponding to eigenvectors
\begin{align}
\ket{\chi_1} &= \frac{1}{\sqrt{2}}(\ket{0} + \ket{1}),  \\
\ket{\chi_{i}} &= \frac{1}{\sqrt{2}}c^\dagger(\ket{0}+i\ket{1}), \\
\ket{\chi_{-1}} &= \frac{1}{\sqrt{2}}(\ket{0} + \ket{1}), \\
\ket{\chi_{-i}} &= \frac{1}{\sqrt{2}}c^\dagger(\ket{0}-i\ket{1}) .
\end{align}
Note that all the eigenvectors have definite fermion parity. We can now construct the projector Hamiltonian for each phase.
\begin{align}
H_1 &= -\frac{1}{2}    ( 1+X ) - \frac{1}{2} (1+(-1)^n), \\
H_{i} &= -\frac{1}{2} ( 1+Y)- \frac{1}{2} (1-(-1)^n), \\
H_{-1} &= -\frac{1}{2} ( 1-X )- \frac{1}{2} (1+(-1)^n) ,\\
H_{-i} &= -\frac{1}{2} ( 1-Y) - \frac{1}{2} (1-(-1)^n).
\end{align}
The unitary that cycles through the four phases is given by
\begin{align}
U & = \ket{\chi_1}\bra{\chi_{-i}} + \ket{\chi_{-i}}\bra{\chi_{-1}} + \ket{\chi_{-1}}\bra{\chi_{i}} + \ket{\chi_i}\bra{\chi_{1}} \nonumber\\
 &= S\gamma
\end{align}
Note that $U^2=Z$ creates the bosonic embedded phase protected by $\mathbb Z_2$. Furthermore, one can also check that $U$ and $\chi$ commute up to a phase factor of $i$, corresponding to the 1D representation of $\ket{\chi_{i}}$ under $\mathbb Z_4^F$.

\subsection{Hamiltonian for the Generating Phase}
Let us now introduce the 1D model. Our system has $2N$ sites with $\mathbb Z_4$ qudits living on the integer sites and $\mathbb Z_4^F$ qubit+fermions living on the half-integer sites. To distinguish the two, the qubit is labeled using, $g=0,1$ while the qudit consists of vectors $\ket{h}$, where $h=0,1,2,3$ modulo 4. A review for the $\mathbb Z_m \times \mathbb Z_m$ decorated domain wall procedure is given in Appendix \ref{ZN}. Here, we concentrate on $m=4$. For qudits, the Pauli matrices are generalized to clock and shift matrices
\begin{align}
\mathcal{Z} \ket{h} &= i^h \ket{h},   & \mathcal{X} \ket{h} &= \ket{h+1}.
\end{align} 
Hence, the symmetries are generated by
\begin{align}
\chi_1 &=  \prod_{i=1}^N \mathcal{X}_{i}, & \chi_2 &= \prod_{i=1}^N X_{i+\frac{1}{2}} (-1)^{n_{i+\frac{1}{2}} g_{i+\frac{1}{2}}},
\end{align}
where $\chi_2^2= P_f$. The bare Hamiltonian is given by the sum of projectors for each site.
\begin{align}
H_0 =&- \sum_i \frac{1}{4} \left (1 + \mathcal{X}_{i} + \mathcal{X}_{i}^2 + \mathcal{X}_{i}^3 \right) \nonumber\\
& - \sum_i \left [ \frac{1}{2} (1+X_{i+\frac{1}{2}}) + \frac{1}{2} \left (1+(-1)^{n_{i+\frac{1}{2}}}\right ) \right].
\label{equ:bareZ4f}
\end{align}
Let us define the controlled gates $C_iZ_l$, $C_iS_l$ and $C_i\gamma_l$  in the following way
\begin{align}
C_iZ_l\ket{h_i,g_l} &= (-1)^{h_ig_l} \ket{h_i,g_l},\\
C_iS_l\ket{h_i,g_l} &= i^{h_ig_l} \ket{h_i,g_l},\\
C_i\gamma_l(\ket{h_i} \otimes \ket{f}) &=  \ket{h_i} \otimes \gamma_l^{h_i} \ket{f}.
\end{align}

Our unitary can then be defined as

\begin{align}
U =& \mathcal{Z}_1^2 \prod_{i=N}^1 C_{i}S_{i+\frac{1}{2}} C_{i}\gamma_{i+\frac{1}{2}} C_{i+1}S_{i+\frac{1}{2}}^\dagger C_{i+1}\gamma_{[i+1]-\frac{1}{2}}^\dagger \nonumber\\
=&\mathcal{Z}_1^2  \left((S_{N+\frac{1}{2}}\gamma_{N+\frac{1}{2}})^{h_{N}} (S_{\frac{1}{2}}\gamma_{\frac{1}{2}})^{-h_1} \right)  \nonumber\\
 & \times (S_{N-\frac{1}{2}}\gamma_{N-\frac{1}{2}})^{h_{N-1} - h_{N}}\cdots (S_{1+\frac{1}{2}}\gamma_{1+\frac{1}{2}})^{h_{1} - h_{2}}
\end{align}
This unitary decorates a non-trivial 0D $\mathbb Z_4^F$ SPT on the $\mathbb Z_4$ domain wall, where the $\mathbb Z_4^F$ charge attached (the number of times $S\gamma$ is applied) is equal to the $\mathbb Z_4$ group element labeling the domain wall.

In the case that fermion parity cannot be factored out from the total symmetry group, one has to use the so-called $\mathcal G$-spin structure instead of the regular spin structure\cite{KapustinTurzilloYou2016}. To avoid this subtlety, we will only refer to boundary conditions of the fermions. Explicitly, the unitaries for the antiperiodic ($\gamma_{\frac{1}{2}}=-\gamma_{N+\frac{1}{2}}$) and periodic ($\gamma_{\frac{1}{2}}=\gamma_{N+\frac{1}{2}}$) boundary conditions are respectively
\begin{align}
U_{AP} =&  (S_{N+\frac{1}{2}}\gamma_{N+\frac{1}{2}})^{h_{N}-h_1} \cdots (S_{1+\frac{1}{2}}\gamma_{1+\frac{1}{2}})^{h_{1} - h_{2}},\\
U_{P} =& \mathcal{Z}_1^2 (S_{N+\frac{1}{2}}\gamma_{N+\frac{1}{2}})^{h_{N}-h_1} \cdots (S_{1+\frac{1}{2}}\gamma_{1+\frac{1}{2}})^{h_{1} - h_{2}}.
\end{align}

First, conjugating by $\chi_1$, we see that the controlled-Majoranas are invariant under $h_i \rightarrow 1+h_i$ (mod 4). Thus we find that they commute for antiperiodic boundary conditions and anticommute for periodic boundary conditions. On the other hand, they commute with $\chi_2$ for both boundary conditions, which also implies that they commute with fermion parity.

Evolving, we find the dressed operators
\begin{align}
\bar{\mathcal{X}}_{i} &= S_{i-\frac{1}{2}}^\dagger \gamma_{i-\frac{1}{2}} \mathcal{Z}_{i}^2 \mathcal{X}_{i} \gamma_{i+\frac{1}{2}} S_{i+\frac{1}{2}} \mathcal{Z}_{i+1}^2\\
\bar X_{i+\frac{1}{2}} &= \mathcal{Z}_{i}  X_{i+\frac{1}{2}} \mathcal{Z}_{i+1}^\dagger C_{i}Z_{i+\frac{1}{2}} C_{i+1}Z_{i+\frac{1}{2}},\\
(-1)^{\bar n_{i+\frac{1}{2}}} &= \mathcal{Z}_{i}^2 (-1)^{ n_{i+\frac{1}{2}}} \mathcal{Z}_{i+1}^2.
\end{align}
In particular, we see that $\bar X_{i+\frac{1}{2}}$ is equal to $\pm X_{i+\frac{1}{2}}, \pm Y_{i+\frac{1}{2}}$ depending on the adjacent qudits. The dressed Hamiltonian can be obtained by replacing the bare operators in Eq. \eqref{equ:bareZ4f} with the dressed ones. We can see that the Hamiltonian is uniform only for the periodic boundary conditions, and $\bar{\mathcal{X}}_1$ obtains a negative coefficient for antiperiodic boundary conditions. 

\subsubsection{Projective Representation of the Boundary}

The dressed operators are all one when acting on the ground state. Therefore, the bare operators act as
\begin{align}
\mathcal{X}_{i} &= - S_{i-\frac{1}{2}} \gamma_{i-\frac{1}{2}} \mathcal{Z}_{i}^2  \gamma_{i+\frac{1}{2}} S_{i+\frac{1}{2}}^\dagger \mathcal{Z}_{i+1}^2,\\
X_{i+\frac{1}{2}} &= \mathcal{Z}_{i}  \mathcal{Z}_{i+1}^\dagger C_{i}Z_{i+\frac{1}{2}} C_{i+1}Z_{i+\frac{1}{2}},\\
(-1)^{ n_{i+\frac{1}{2}}} &= \mathcal{Z}_{i}^2 \mathcal{Z}_{i+1}^2.
\end{align}
In particular, the onsite symmetry generators of $\chi_2$ are found to be
\begin{align}
X_{i+\frac{1}{2}} (-1)^{  n_{i+\frac{1}{2}}h_{i+\frac{1}{2}}} &= \mathcal{Z}_{i}  \mathcal{Z}_{i+1}^\dagger C_{i}Z_{i+\frac{1}{2}} C_{i+1}Z_{i+\frac{1}{2}}  \mathcal{Z}_{i}^{2h_i} \mathcal{Z}_{i+1}^{2h_i} \nonumber\\
&= \mathcal{Z}_{i}  \mathcal{Z}_{i+1}^\dagger.
\end{align}
Hence, we find that the symmetry operators on the left edge are
\begin{align}
\chi_1^L &= \mathcal{X}_1  S_{1+\frac{1}{2}} \gamma_{1+\frac{1}{2}} \mathcal{Z}_{2}^2, \\
\chi_2^L &= X_{1+\frac{1}{2}} (-1)^{n_{1+\frac{1}{2}}h_{1+\frac{1}{2}}} \mathcal{Z}_{2},
\end{align}
which implies
\begin{align}
\chi_1^L \chi_2^L &= -i \chi_2^L \chi_1^L.
\end{align}
This is the projective representation of $\mathbb Z_4 \times \mathbb Z_4^F$ and proves that our model gives the generating phase.

\subsubsection{Squaring Gives the Bosonic-Embedded Phase}
If we conjugate the operators using $U^2$, we find that the dressed operators are
\begin{align}
\bar{\mathcal{X}}_i &= - \mathcal{Z}_{i-1}^2 Z_{i-\frac{1}{2}}  \mathcal{X}_i Z_{i+\frac{1}{2}} \mathcal{Z}_{i+1}^2,\\
\bar X_{i+\frac{1}{2}} &= \mathcal{Z}_{i}^2  X_{i+\frac{1}{2}} \mathcal{Z}_{i+1}^2,\\
(-1)^{\bar n_{i+\frac{1}{2}}} &= (-1)^{ n_{i+\frac{1}{2}}}.
\end{align}
The corresponding Hamiltonian without fermions is the bosonic SPT protected by $\mathbb Z_4 \times \mathbb Z_2$.

\subsubsection{FDLU}
Similarly to the previous section, the terms in $U$ do not commute and so we need to rewrite $U$ as an FDLU. We again introduce ancilla fermions living at positions $i$ with Majorana operators $\eta$ and $\tilde \eta$. One can then rewrite the unitary as

\begin{align}
U =&  \prod_i \left[\tilde \eta_{i}^{h_{i}} \eta_{i}^{h_{i}} \right ]  \prod_i \left [ \tilde \eta_{i+1}^{h_{i+1}}\gamma_{i+\frac{1}{2}}^{h_{i}} \gamma_{[i+1]-\frac{1}{2}}^{h_{i+1}} \eta_{i}^{h_{i}} \right] \nonumber\\
 &\prod_{i} \left[ C_{i}S_{i+\frac{1}{2}}C_{i+1}S_{i+\frac{1}{2}}^\dagger \right ],
\end{align}
which is a three-layer FDLU that is symmetric with respect to $P_f$. Writing it out explicitly for the two boundary conditions, we have
\begin{align}
U_{AP} = &\mathcal Z_1^2  \prod_i \left[\tilde \eta_{i}^{h_{i}} \eta_{i}^{h_{i}} \right ]  \prod_i \left [ \tilde \eta_{i+1}^{h_{i+1}}\gamma_{i+\frac{1}{2}}^{h_{i}+h_{i+1}} \eta_{i}^{h_{i}} \right] \nonumber\\
 &\prod_{i} \left[C_{i}S_{i+\frac{1}{2}}C_{i+1}S_{i+\frac{1}{2}}^\dagger\right ],\\
 U_{P} =&   \prod_i \left[\tilde \eta_{i}^{h_{i}} \eta_{i}^{h_{i}} \right ]  \prod_i \left [ \tilde \eta_{i+1}^{h_{i+1}}\gamma_{i+\frac{1}{2}}^{h_{i}+h_{i+1}} \eta_{i}^{h_{i}} \right] \nonumber\\
 &\prod_{i} \left[C_{i}S_{i+\frac{1}{2}}C_{i+1}S_{i+\frac{1}{2}}^\dagger\right ].
\end{align}

\section{Model for 2D SPT with $G=\mathbb Z_2 \times \mathbb Z_2^F$}
\label{2D}
\begin{figure}
\centering
\includegraphics{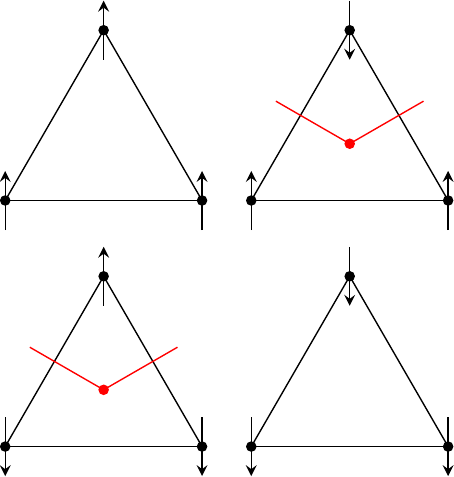}
\caption{The fermion at the center of a triangle is excited (red dot) along the corners of the domain wall (red line) in the dual lattice}
\label{fig:DDW2D}
\end{figure}

\begin{figure}
\includegraphics{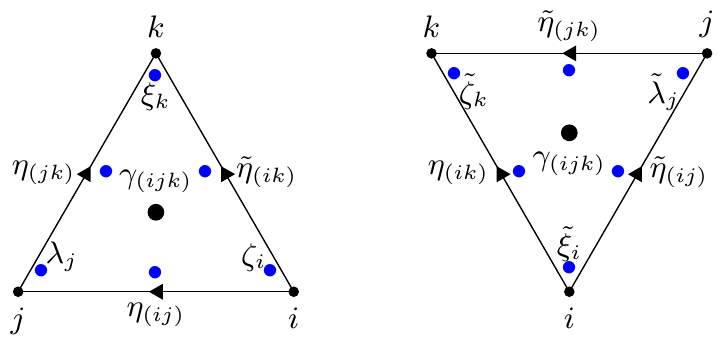}
\caption{Local fermionic Hilbert space within $\Delta$ and $\nabla$ type triangles on the triangular lattice. The edge has one fermion site split into two Majorana fermions $(\eta/\tilde \eta)$ and the corner has three fermion sites split into six Majoranas $(\zeta/\tilde  \zeta, \lambda/\tilde  \lambda, \xi/\tilde  \xi)$ shown in blue. The Majorana operators are paired up across the edges/corners and squared away in the unitary.}
\label{fig:Gamma}
\end{figure}

We now turn our attention to two-dimensional SPTs protected by $G=\mathbb Z_2 \times \mathbb Z_2^F$. In the free case, they are classified by $\mathbb Z$ with topological index $\nu$ generated by stacking $\nu$ copies of $p+ ip$ superconductors and $\nu$ copies of $p-ip$ superconductors, where the latter is charged under the $\mathbb Z_2$ symmetry. Similarly to the topological superconductors in 1D, they reduce to $\mathbb Z_8$ under interactions\cite{GuLevin2014,Kapustinetal2015}, and the $\nu=4$ can be created by embedding the Levin-Gu model\cite{LevinGu2012} (reviewed in Appendix \ref{LevinGu}). Again, only the even $\nu$'s are captured by supercohomology and in this section, we will construct a commuting projector model for the $\nu=2$ phase.

 Our model is defined on a triangular lattice with qubits on the vertices and fermions at the center of the triangles $(ijk)$, which come in two types: $\Delta$ and $\nabla$. Following Eq. \eqref{equ:bareH}, our bare Hamiltonian is
\begin{equation}
H_0=-\sum_i \frac{1}{2}(1+X_i) -\sum_{\Delta,\nabla}\frac{1}{2}\left(1+(-1)^{n_{(ijk)}}\right).
\label{equ:bare2D}
\end{equation}
\subsection{Model and Ground State}
To emulate the decorated domain wall procedure that we have done in 1D, recall that the unitary in Eq. \eqref{equ:U_1D} was written as a product of  controlled-Majorana gates $\gamma^{g_i+g_j}$  for each fermions site between two qubits $g_i$ and $g_j$. We can make it more symmetric by rewriting it as $\gamma^{1+g_ig_j+(1-g_i)(1-g_j)}$. The natural generalization to two dimensions is that for each triangle, we apply the controlled-Majorana
\begin{equation}\gamma^{1+g_ig_jg_k+(1-g_i)(1-g_j)(1-g_k)} = \gamma^{g_i g_j + g_j g_k + g_i g_k+ g_i+g_j+g_k}.
\end{equation}
This corresponds to exciting a fermion whenever there are both up and down spins on the surrounding triangle. Alternatively, we can say that a fermion is decorated on every corner of the domain wall defined on the dual lattice as illustrated in Figure \ref{fig:DDW2D}. Nevertheless, we face a problem in 2D because there is no canonical way to order the fermions.

In order to write down an FDLU such that each gate is fermion parity even, we first need to enlarge our Hilbert space by including the following:\\
1. A fermion site at every edge corresponding to Majorana operators $\eta/\tilde \eta$ at each side of the triangle.\\
2. Three fermion sites at each vertex corresponding to six Majorana operators  $\zeta/\tilde  \zeta, \lambda/\tilde  \lambda, \xi/\tilde  \xi$ at each corner of the six triangles containing that vertex. This is shown in Figure \ref{fig:Gamma}. We now introduce the following gates.

\begin{align}
\Gamma^\Delta_{ijk} &= \gamma_{(ijk)}^{g_i g_j + g_j g_k + g_i g_k+ g_i+g_j+g_k} \eta_{(ij)}^{g_i g_j}  \eta_{(jk)}^{g_j g_k} \tilde \eta_{(ik)}^{g_i g_k} \zeta_i^{g_i} \lambda_j^{g_j} \xi_k^{g_k},\\
\Gamma^\nabla_{ijk} &=\tilde \xi_k^{g_k} \tilde  \lambda_j^{g_j} \tilde \zeta_i^{g_i} \eta_{(ik)}^{g_i g_k} \tilde  \eta_{(jk)}^{g_j g_k} \tilde \eta_{(ij)}^{g_i g_j}   \gamma_{(ijk)}^{g_i g_j + g_j g_k + g_i g_k+ g_i+g_j+g_k}.
\end{align}

These gates act on each $\Delta/\nabla$ triangle and commute with fermion parity. Similarly to the 1D unitary, we must square all the ancilla Majoranas out in pairs so that we are only left with fermions excited along the domain wall corners. However, this is still not enough to make $U$ overall symmetric. It turns out that to fix this, we must also include a $CC\Ri/CC\Ric$ operator for every $\Delta/\nabla$ triangle respectively. This phase of $\pm i$ assigned can in a way be thought of as the cochain $\omega(1,1,1) =i$ in the supercohomology data. To conclude, our unitary operator is
\begin{align}
U =& \prod_i  \zeta_i^{g_i} \tilde \zeta_i^{g_i} \lambda_i^{g_i} \tilde  \lambda_i^{g_i} \xi_i^{g_i} \tilde \xi_i^{g_i} \prod_{<ij>} \eta_{(ij)}^{g_ig_j} \tilde \eta_{(ij)}^{g_ig_j} \nonumber\\
 & \prod_{\Delta}  CC\Ri_{ijk} \Gamma^\Delta_{ijk} \prod_{\nabla} CC\Ric_{ijk} \Gamma^\nabla_{ijk}.
 \label{equ:U2}
\end{align}
Hence, we have written $U$ as an FDLU with three layers. We remark that in principle, we can similarly square out all the ancilla Majoranas and get some spin-dependent phase, but such term is very complicated and also depends on our ordering of the remaining Majoranas at the center of triangles.

Conjugating our bare Hamiltonian with this unitary operator through a similar procedure, we obtain
\begin{figure}
\includegraphics{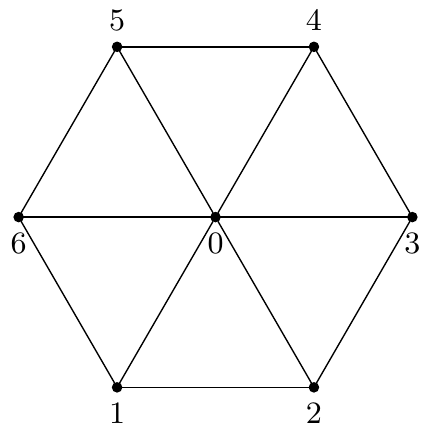}
\caption{Our choice of ordering the nearest-neighbor sites for each dressed qubit term in the Hamiltonian.}
\label{fig:ordering}
\end{figure}
\begin{widetext}
\begin{align}
H_2=-\sum_i & \frac{1}{2}\bigg [1+Z_0X_0 \gamma_{(012)}^{g_1 +g_2+1}  \gamma_{(023)}^{g_2 +g_3+1} \gamma_{(034)}^{g_3 +g_4+1}  \gamma_{(045)}^{g_4 +g_5+1} \gamma_{(056)}^{g_5 +g_6+1}  \gamma_{(061)}^{g_6 +g_1+1}  C\Ri_{12} C\Ric_{23} C\Ri_{34} C\Ric_{45} C\Ri_{56} C\Ric_{61}  \nonumber \\
 &  CZ_{01}  CZ_{02} CZ_{03} CZ_{04} CZ_{05} CZ_{06} CZ_{12} CZ_{23} CZ_{24}CZ_{25} CZ_{26} CZ_{35}  CZ_{36} CZ_{45} CZ_{46} \bigg] \nonumber\\
 -\sum_{\Delta,\nabla} & \frac{1}{2} \left [1+(-1)^{n_{i+\frac{1}{2}}}Z_iZ_jZ_kCZ_{ij}CZ_{jk}CZ_{ik}\right].
 \label{equ:H_2}
 \end{align}
\end{widetext}
Here, we labeled each qubit $i$ as site $0$ and denoted the nearest-neighbor qubits as sites 1-6 in the order shown in Figure \ref{fig:ordering}. This ordering is arbitrary, and as a result, a different ordering will require us to reorder the $C\gamma$ operators in the dressed qubit, which gives a different string of $CZ$ operators, but note that the product $CZ_{0i}$ where $i=1,...,6$ is always present.

We explicitly check in Appendix \ref{properties} that the following properties are satisfied:\\
\ref{Projector} All terms in the Hamiltonian are projectors.\\
\ref{Hermitian} The dressed qubit projector is Hermitian.\\
\ref{SymmetricProjector} The projectors commute with the symmetries.\\
\ref{SymmetricU} $U$ is an overall symmetric FDLU.\\
\ref{U2givesnu4} Conjugating $U^2$ to the Hamiltonian gives the Levin-Gu Hamiltonian.

The methods used, although more arduous, is in essence the same as the 1D model. The property \ref{SymmetricU} implies that the Hamiltonian must realize one of the eight SPT phases, while the property \ref{U2givesnu4} implies that the unitary gives either the $\nu = 2$ or $\nu = 6$ phase. Here, we have chosen to label this model with $\nu=2$. Nevertheless, we can also construct the $\nu=6$ phase using a similar unitary. Recall that the $\nu=6$ phase can be obtained by reflecting the $\nu=2$ phase. This swaps the orientations of the $\Delta$ and $\nabla$ triangles, and we see that we only need to swap the assignments of $CC \Ri$ and $CC\Ric$ in our unitary. This can be thought of as $\omega(1,1,1)=-i$ in the supercohomology data.

\subsection{Ground State Wave Function}

\begin{figure}
\includegraphics{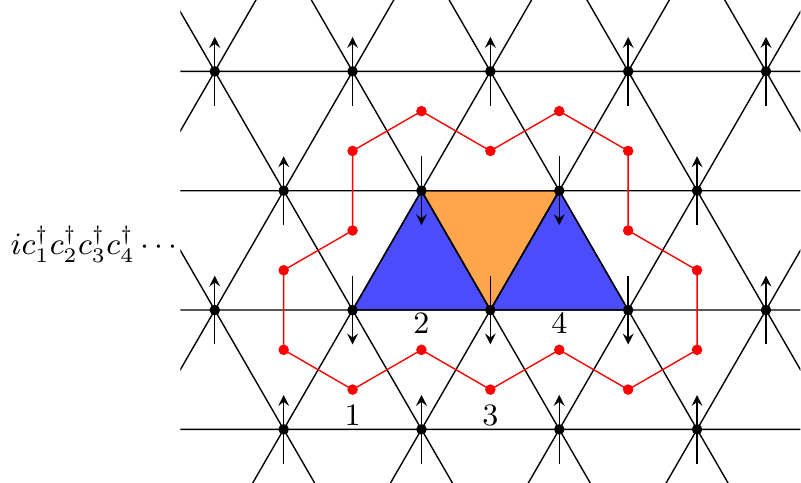}
\caption{Fermions are excited along the corners of each domain wall configuration written starting from a fermion at the center of a $\Delta$ triangle and proceeding counterclockwise. There is also a phase factor of $i$ or $-i$ for any $\Delta$ or $\nabla$ triangle in the down spin region shown respectively in blue and orange. In this configuration, the total phase factor is $i^2 \cdot (-i) = i$. The ground state is a superposition of all such configurations.}
\label{fig:GS2D}
\end{figure}

To find the ground state, we can start with a state with all spins pointing up and act with each of the dressed qubit terms from the Hamiltonian. This flips a spin and decorates fermions around it in a particular manner. Doing so, we find that the ground state is a superposition of all domain wall configurations with a factor of $\pm i$ for each $\Delta/\nabla$ triangle of down spin regions and fermions excited at every corner along the domain walls. This is illustrated in Figure \ref{fig:GS2D}. Since we must choose an order for the fermions around the domain wall, we choose the following convention. Pick a $\Delta$ triangle on the domain wall loop and write the creation operators from left to right in a counterclockwise manner. We can see that this will not depend on the $\Delta$ triangle we choose. Moreover, since the qubit terms commute with the symmetry, our ground state wave function is symmetric by construction.

There are also two nice properties to note about this wave function in relation to the other even $\nu$ phases . First, upon reflection, we must swap the assignment of $i$ and $-i$ to $\nabla$ and $\Delta$ triangles enclosed by the down spins, which produces the ground state for $\nu=6$. Second, by applying $U$ twice, all the fermions around the domain walls are squared away. One can show that the sign from commuting all the Majoranas around a domain wall loop is always equal to the vertices minus edges inside that domain wall modulo 2. Furthermore, one gets a minus sign for each triangle (face) enclosed by the down spins from squaring $\pm i$. Consequently, the wave function gets a sign according to the Euler characteristic of the down spins modulo 2, which in turn is equivalent to a minus sign for every domain wall loop. This is exactly the Levin-Gu ground state (a review of this ground state is given in Appendix \ref{LevinGu}).

\subsection{Edge properties}
\begin{figure}
\includegraphics{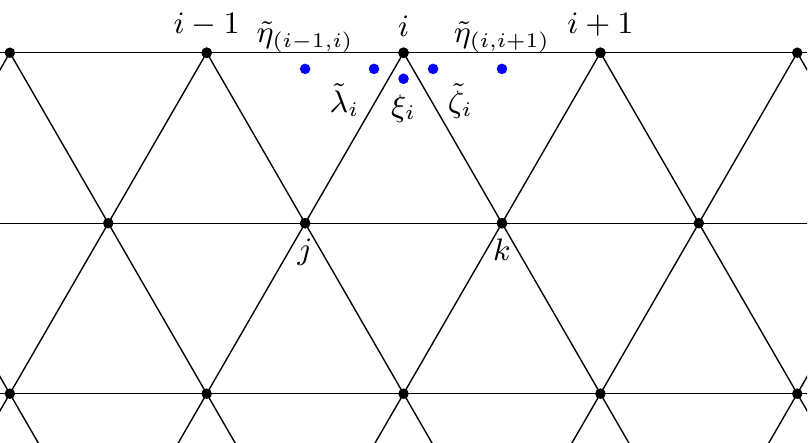}
\caption{Edge of a triangular lattice}
\label{fig:edge}
\end{figure}
We will now analyze the edge of this SPT and show that it has a non-trivial boundary. We do so by analyzing how the qubits transform near the edge\cite{LevinGu2012,Yoshida2016}.

At the boundary, the qubits are dressed differently from those in the bulk because only part of the unitary is applied. Hence, the ancilla fermions $\eta$ and $\tilde \eta$ that were initially squared away on a closed manifold will now excite fermions near the boundary. Consider the qubit at site $i$ on the boundary shown in Figure \ref{fig:edge}. Upon conjugating, we find the dressed qubits on the boundary to be

\begin{align}
\label{Xbarboundary}
\bar X_i &= X_i Z_i C\Ric_{i-1,j}  C\Ric_{j,k}  C\Ric_{j,i+1} CZ_{i,j} CZ_{i,k}\tilde \eta_{(i-1,i)}^{g_{i-1}}  \nonumber\\
& \gamma_{(i-1,j,i)}^{g_{i-1}+g_{j}+1} \tilde\lambda_i \xi_i \tilde\zeta_i \gamma_{(j,i,k)}^{g_{j}+ g_{k}+1} \gamma_{(i,k,i+1)}^{g_{k} +g_{i+1}+1}  \tilde \eta_{(i,i+1)}^{g_{i+1}}.
\end{align}

On the other hand, the dressed $Z, \Ri$ and Majorana operators on the boundary are the same as the bare ones. These dressed operators still obey the Pauli algebra. Moreover, they commute with the dressed operators in the bulk because all the gates in $U$ commute with one another. 

The symmetry leaves the dressed spins invariant in the bulk, but act non-trivially on the edge. Upon conjugating with the $\mathbb Z_2$ symmetry $\chi$, we find
\begin{align}
\chi \bar X_i \chi &=  - i \Ri_{i-1} \eta_{i-\frac{1}{2}} \bar X_i \eta_{i+\frac{1}{2}} \Ri_{i+1},\\
\chi \bar Z_i \chi &= -\bar Z_i,
\end{align}
where we have renamed the Majorana operators $\tilde \eta_{(i-1,i)} = \eta_{i-\frac{1}{2}}$ and $\tilde \eta_{(i,i+1)} = \eta_{i+\frac{1}{2}}$. At first glance, the expression $\chi \bar X \chi$ does not look Hermitian. However, this is because $\bar X$ is dressed with fermions, and so from Eq. \eqref{Xbarboundary}, we see that $\bar X$ has an interesting commutation relation with the Majoranas,
\begin{align}
\eta_{i-\frac{1}{2}} \bar X_i &= \bar X_i Z_{i-1} \eta_{i-\frac{1}{2}},\\
\eta_{i+\frac{1}{2}} \bar X_i &= \bar X_i Z_{i+1} \eta_{i+\frac{1}{2}}.
\end{align}
Let us consider the following edge Hamiltonian
\begin{align}
H_\text{edge} =& -\sum_i h_i \left ( \bar X_i -i \bar \Ri_{i-1} \eta_{i-\frac{1}{2}} \bar X_i \eta_{i+\frac{1}{2}} \bar \Ri_{i+1} \right)\nonumber\\
 &-\sum_i h_{i+\frac{1}{2}} (-1)^{n_{i+\frac{1}{2}}}.
\end{align}
where the coefficients $h$ are all real. It can be seen that the Hamiltonian respects the global $\mathbb Z_2$ symmetry and fermion parity. 

We will show that this Hamiltonian cannot be both symmetric and gapped at the same time. To do so, we notice that the Hamiltonian respects the following $\bar {\mathbb  Z}_2$ symmetry,
\begin{equation}
 \bar \chi =\prod_i \bar X_i,
 \end{equation}
where $\bar X_i$ is the dressed $X_i$ on the edge in Eq. \eqref{Xbarboundary}. This symmetry is denoted a bar to emphasize that it is a different symmetry from the action of the global $\mathbb Z_2$ symmetry on the edge.
The Hamiltonian is also invariant under fermion parity restricted to the edge. Therefore, we can view this Hamiltonian as a system with $\bar {\mathbb Z}_2 \times \mathbb Z_2^F$ symmetry.
To argue that this Hamiltonian is gapless, let us split the Hamiltonian into two parts
\begin{align}
H_0 &= -\sum_i h_i \bar X_i  +  \frac{1}{2}h_{i+\frac{1}{2}} (-1)^{n_{i+\frac{1}{2}}}\\
H_1 &= -\sum_i h_i \left ( -i \bar \Ri_{i-1} \eta_{i-\frac{1}{2}} \bar X_i \eta_{i+\frac{1}{2}} \bar\Ri_{i+1} \right) + \frac{1}{2} h_{i+\frac{1}{2}} (-1)^{n_{i+\frac{1}{2}}}.
\label{equ:edge}
\end{align}
$H_0$ is just the Hamiltonian of a trivial SPT since the ground state is a product state. On the other hand, we can show that $H_1$ is the Hamiltonian of a supercohomology phase SPT protected by $\bar{\mathbb  Z}_2 \times \mathbb Z_2^F$ (i.e., the non-trivial phase without Majorana edge modes). To prove the non-trivialness of the latter, we can compute the topological invariant\cite{Tantivasadakarn2017}
\begin{equation}
\frac{_R\bra{\psi}\bar \chi\ket{\psi}_{R}}{_{NS}\bra{\psi}\bar \chi\ket{\psi}_{NS}},
\end{equation}
which is 1 for the trivial phase, but $-1$ for a non-trivial phase (see Appendix \ref{topinv} for a discussion of this invariant). For our Hamiltonian, we see that the only term in $H_1$ that changes sign when we change boundary conditions from $\eta_{1-1/2}=\eta_{N-1/2}$ to $\eta_{1-1/2}=-\eta_{N-1/2}$ is
\begin{equation}
-i \bar \Ri_{N-1} \eta_{N-\frac{1}{2}} \bar X_N \eta_{1-\frac{1}{2}} \bar\Ri_{1}.
\end{equation}
As a result, the Hamiltonians with different boundary conditions are related by a conjugation of $\bar Z_N$, and we can therefore relate the two ground state wave functions via
\begin{equation}
\ket{\psi}_{R} = \bar Z_N \ket{\psi}_{NS}.
\end{equation}
Therefore, the ground state of $H_1$ satisfies,
 \begin{equation}
 _R\bra{\psi}\chi\ket{\psi}_{R} = _{NS}\bra{\psi} Z_N\bar \chi Z_N \ket{\psi}_{NS} = - _{NS}\bra{\psi} \bar \chi \ket{\psi}_{NS},
 \end{equation}
showing that $H_1$ realizes the non-trivial SPT.

Having proved that $H_0$ and $H_1$ are in two different SPT phases under this symmetry, we further notice that the action of the global $\mathbb Z_2$ symmetry on the boundary change $H_0$ to $H_1$ and vice versa. Therefore, the Hamiltonian $H_\text{edge}=H_0+H_1$ is at the phase transition between the two SPT phases, proving that it cannot be both symmetric and gapped.

We remark that for the triangular lattice, there is a second type of edge (zigzag). Repeating the calculation on this edge, we find a different edge Hamiltonian:
\begin{align}
H_\text{edge} = -\sum_i \left ( \bar X_i + \bar \Ri_{i-1} \eta_{i-\frac{1}{2}} \bar X_i \bar Z_i \eta_{i+\frac{1}{2}} \bar \Ri_{i+1} \right) + (-1)^{n_{i+\frac{1}{2}}},
\end{align}
An identical analysis show that this Hamiltonian also cannot be both symmetric and gapped.

\section{Towards a General Construction for Finite Abelian Unitary Symmetry Groups}
\label{general}
In this section, we compare our results to the K\"unneth formula in supercohomology and show that they are consistent if we consider more general symmetry groups. For this discussion, we limit ourselves to symmetry groups of the form $G=\mathbb Z_2^F \times \prod_i \mathbb Z_{N_i}$.

In one dimension, we have the choice of exciting the fermion along the domain wall of each $\mathbb Z_{N_i}$. This can only be done consistently if $N_i$ is even and so the possible ways to do this are consistent with the K\"unneth formula for $\mathcal H^1(G,\mathbb{Z}_2)$ from supercohomology:
\begin{align}
\mathcal H^1(G,\mathbb{Z}_2) \cong \prod_i \mathbb{Z}_{(N_{i},2)},
\end{align}
where $(..,..)$ denotes the GCD.

\begin{figure}
\includegraphics{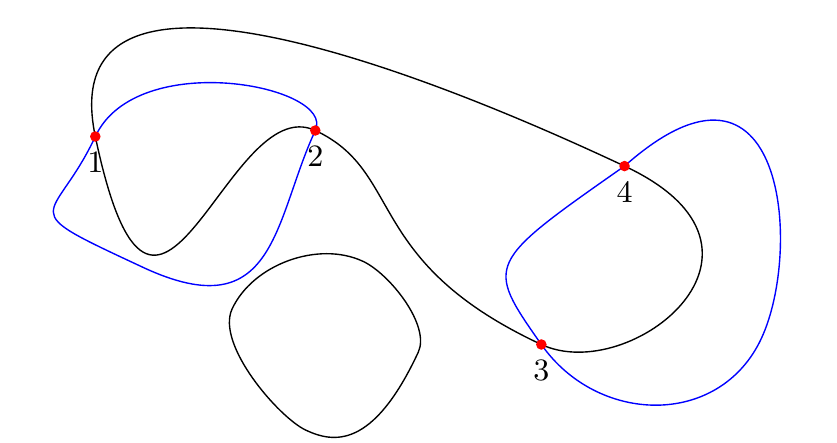}
\caption{Schematic ground state wave function of the $\mathbb Z_2 \times \mathbb Z_2 \times \mathbb Z_2^F$ SPT. For each domain wall configuration shown in black and blue, fermions are excited in pairs (1, 2 and 3, 4) at the intersection of the domain walls. This is the square root of the bosonic phase, which assigns a minus sign to each pair of domain wall intersections.}
\label{fig:intersection}
\end{figure}

In two dimensions, for each $\mathbb Z_{N_i}$ we can similarly excite the fermion at the corners of the domain wall if $N_i$ is even. However, there is another possibility when there are two or more domain walls. As an example, consider the symmetry $G=\mathbb Z_2 \times \mathbb Z_2 \times \mathbb Z_2^F$. Our product state is a superposition of the domain walls of the two $\mathbb Z_2$ symmetries. We can then find an FDLU that excites fermions at every pair of domain wall intersections (up to minus signs) as shown in Figure \ref{fig:intersection}. This can also be viewed as decorating a supercohomology $\mathbb Z_2 \times \mathbb Z_2^F$ 1D SPT on the domain wall of the other $\mathbb Z_2$ symmetry. This wave function is invariant under both symmetries and fermion parity. Furthermore, by applying the unitary twice, we can square away each pair of fermions and get a minus sign for every pair of domain walls intersections. This is just the wave function for the $\mathbb Z_2 \times \mathbb Z_2$ bosonic SPT \footnote{specifically, the phase represented by the cocycle $a_1 \cup da_2$}, and so the fermionic phase constructed is the square root of this bosonic SPT.

For a general finite Abelian group, the K\"unneth formula for $\mathcal H^2(G,\mathbb{Z}_2)$ is
\begin{align}
\mathcal H^2(G,\mathbb{Z}_2) \cong \prod_i \mathbb{Z}_{(N_{i},2)}\prod_{i<j} \mathbb{Z}_{(N_i,N_j,2)}.
\end{align}
Since the fermions are 0D objects, they can only be decorated on intersections of at most two 1D domain walls. Thus, one should be able to create the supercohomology phases in 2D with this symmetry by either decorating fermions at the corners of one symmetry domain wall, or at the intersection of two symmetry domain walls. This is consistent with the above formula.

In 3D, the K\"unneth formula for $\mathcal H^3(G,\mathbb{Z}_2)$ is
\begin{align}
\mathcal H^3(G,\mathbb{Z}_2) \cong \prod_{i} \mathbb Z_{(N_{i},2)} \prod_{i<j} \mathbb Z_{(N_i,N_j,2)}^2 \prod_{i<j<k} \mathbb Z_{(N_i,N_j,N_k,2)}.
\end{align}
We conjecture that the first product corresponds to decorating fermions on the corners of one symmetry domain wall. The second product should correspond to decorating a fermionic 2D SPT with one symmetry on the other symmetry's domain wall and vice versa. Finally, the last product should correspond to decorating fermions on the intersection of three symmetry domain walls. These 3D constructions are also consistent with a closely related process of dimensional reduction for supercohomology SPTs from 3D to 2D\cite{WangLevin2014,WangLevin2015,Tantivasadakarn2017}.

Our speculation shows that it should be possible to construct full commuting projectors for other supercohomology phases with finite Abelian unitary symmetries using fermionic decorated domain walls. Indeed, if such construction could be done in general, it would imply that the fermionic decorated domain wall realizes of the K\"unneth formula for $\mathcal H^{d}(G,\mathbb Z_2)$ in group supercohomology, in the same way that the bosonic decorated domain walls is the physical interpretation of the K\"unneth formula for $\mathcal H^{d+1}(G,U(1))$ in group cohomology. We leave the verification of this hypothesis to future work.

At the same time, we should be able to construct SPTs for other Abelian symmetry groups outside the considerations of supercohomology, such as decorating $\mathbb Z_{4}^F$ (or in general $\mathbb Z_{2m}^F$ where $m \in \mathbb Z^+$) on the domain walls of other symmetries. These phases have so far only been classified in special cases using for example, the anomalous action on the boundary\cite{ElseNayak2014} or from the possible braiding statistics of the gauged theories\cite{Wang2016,WangLinGu2017,ChengTantivasadakarnWang2018}. Some of these symmetries have been shown to admit intrinsically interacting phases that cannot be obtained from free theories. For example, one should be able to construct a $\mathbb Z_{4} \times \mathbb Z_{4} \times \mathbb Z_{4}^F$ SPT in 2D by decorating the $\mathbb Z_{4}^F$ 0D SPT at the intersection of the two $\mathbb Z_{4}$ domain walls, or in other words, putting the $\mathbb Z_{4} \times \mathbb Z_{4}^F$ SPT we constructed in 1D on the domain wall of the other $\mathbb Z_{4}$ symmetry. Such phase is known to be intrinsically interacting\cite{WangLinGu2017}. Furthermore, the supercohomology phase with $\mathbb Z_{2} \times \mathbb Z_{4} \times \mathbb Z_{2}^F$ in 3D is also intrinsically interacting\cite{ChengTantivasadakarnWang2018}. It would be interesting to explore the boundary properties of such models\cite{FidkowskiVishwanathMetlitski2018}. Such constructions and analyses are interesting paths to extend on.

\section{Full Commuting Projectors \& Many- Body Localization}
\label{MBL}
In the previous sections, our Hamiltonians were not yet full commuting projectors because we did not write down the projectors for the excited states. In this section, we will restore them to obtain the FCPH and argue that all the excited states in our models are in the same phase so that the topological properties can be assigned to the FCPH as a whole. We will then discuss about the implication of these FCPHs to Many-Body Localization (MBL) of the corresponding SPT phases.
\subsection{Full Commuting Projectors \& Excited States}
For simplicity, let us start the discussion with our bare Hamiltonian Eq.\eqref{equ:bareH}. To make this into an FCPH, we simply need to restore the projectors of the excited states. 
\begin{align}
H_0 =& - \sum_{i \in V} \left (h_i^+ \frac{1}{2}(1+X_i) \right. &&+  \left.h_i^- \frac{1}{2}(1-X_i) \right )\nonumber\\
 &- \sum_{l \in V^*} \left (h_{l}^0 \frac{1}{2}(1+(-1)^{n_{l}})\right. &&+  \left.h_{l}^1 \frac{1}{2}(1+(-1)^{n_{l}}) \right),
\label{equ:bareFCPH}
\end{align}
where we demand $h_i^- < h_i^+$ and $h_{l}^1 < h_{l}^0$ to preserve the ground state:  $\ket{+}$ states at all vertices and unoccupied fermions at all dual vertices. The bare excitations correspond to $\ket{-}$ states at each vertex and occupying the fermions at each dual vertex. The excited states can be obtained from the ground state of Eq. \eqref{equ:bareH} by applying $Z$ and $\gamma$ to the corresponding vertices and dual vertices we wish to excite, respectively. Since both operators anticommute with the corresponding symmetries, we have shown that all the excited states can be connected to the ground state via a symmetric FDLU. Hence, all of the eigenstates of the FCPH \eqref{equ:bareFCPH} are in the trivial phase.

Now, let us evolve the Hamiltonian \eqref{equ:bareFCPH} with the FDLU $U$ to get the dressed Hamiltonian. We see that the dressed excited states are still related to the dressed ground state by applying $UZU^\dagger$ and $U \gamma U^\dagger$ to the corresponding vertices and dual vertices respectively. Using the fact that $U$ is an overall symmetric FDLU, these operators are still local and anticommute with the corresponding symmetries. Thus, the dressed excited states can be connected to the dressed ground state and therefore are still in the same phase. Hence, we are able to assign the phase to our FCPH. A similar argument can be made for the 1D $\mathbb Z_4 \times \mathbb Z_4^F$ phase.

\subsection{Many-Body Localization}
An application of our FCPHs is towards understanding the Many-Body Localization (MBL) of SPT phases. An isolated quantum system may fail to thermalize when the system has a full set of (quasi)-local integrals of motion, called $l$-bits\cite{Huseetal2014,Serbynetal2013,Swingle2013,Chandranetal2015,Imbrie2016-1,Imbrie2016-2}. For our FCPH, these are the commuting projectors. Therefore, by introducing disorder to the coefficients $h$, we can drive the system to Many-Body localization. 

For SPT phases out of equilibrium, there is also another condition we must check. That is, the topological properties must still hold for excited states for our Hamiltonian so that upon introducing disorder and taking the thermodynamic limit, the system still has the topological properties such as anomalous edge states\cite{Huseetal2013,Bahrietal2015, Chandranetal2014,PotterVishwanath2015,Slagleetal2015}. We have confirmed so in the previous subsection and so our FCPH provides evidence that Many-Body Localization is possible for all of our example SPTs.

We note that while the existence of full commuting projectors provides a necessary condition for MBL, they are not a sufficient condition. For example, if the symmetry of our Hamiltonian is non-Abelian, multiple eigenstates become degenerate triggering resonances which are expected to destroy MBL\cite{PotterVasseur2016}. Fortunately, all the examples we considered have an Abelian symmetry group. Some authors have argued that MBL in dimensions greater than one may be ultimately unstable\cite{DeRoeckHuveneers2017}, although the time scales and system sizes required to check this conjecture remain out of reach, and the conjecture, even if true, may not be of practical relevance. Caveats to this argument have also been advanced\cite{Potirnicheetal2018}.

Comparing to the non-interacting case, we also remark that there are no localized symmetric Wannier functions for free fermions SPTs in higher than one dimension protected by internal symmetries\cite{SoluyanovVanderbilt2011}. This shows that interacting models in the strongly interacting regime can potentially have drastically different physics from their free counterparts.

\section{Discussion}
We have demonstrated how to generalize the Decorated Domain Wall approach from bosons to fermions. In particular, we constructed FDLUs that attach 0D  SPTs ($\mathbb Z_2^F$ or $\mathbb Z_4^F$ charges) to the domain walls or the domain wall corners of the remaining symmetries. Then, using that unitary to evolve a bare commuting projector Hamiltonian, we obtain FCPHs that realizes that particular phase. The commutation relation between the symmetry and the unitary depends on our choice of spin structure. Furthermore, we found that when the unitary operators are applied twice, we obtain a bosonic commuting projector stacked with a trivial fermions independent of the choice of spin structure. This showed that our models are indeed square roots of the bosonic SPT phases. We further derived the anomalous action of the symmetry on the boundary of our three models and obtained the corresponding edge states. Lastly, we argued the possibility of constructing such FCPHs for general symmetries in physical spatial dimensions of interest and that these models admit Many-Body Localized realizations when the symmetry group is Abelian.

One of the downsides of our 2D model is that it is specific to the triangular lattice.  Because we integrated out the ancilla fermions at the corners of the triangles in pairs, we cannot define it for arbitrary triangulations of a 2D surface. Furthermore, in order to get the Levin-Gu model upon squaring, the number of anticommutations of the fermions along the domain wall had to be equal to the number of vertices minus edges inside the domain wall modulo two. This holds for a triangular lattice, but does not work for, e.g., a square lattice. As a result, there is no obvious continuum version of this model.

There are multiple interesting directions for future work. The first is to show whether it is also possible to construct FCPHs for beyond (super)cohomology such as phases in the generalized cohomology classifications. In particular, one could ask whether such a model exists for the $\nu=1$ phase in 2D, or the time--reversal ``$E_8$" phase in 3D\cite{VishwanathSenthil2013,Burnelletal2014}. For the former, the concept of decorated domain walls using higher dimensional fermionic invertible phases has indeed been established and studied\cite{TarantinoFidkowski2016,Wareetal2016,WangNingChen2017,WangGu2018}, but the FDLUs that realize such procedures are still lacking. One can also try to construct models for the intrinsically interacting fermionic SPTs in 2 and 3 dimensions mentioned earlier.

Second, since we have constructed FCPHs for supercohomology phases, they imply that these phases can also be faithfully represented by tensor networks. There has been recent progress in 1D \cite{Bultincketal2017,KapustinTurzilloYou2016,TurzilloYou2017}, and so it should be possible to generalize such results to construct tensor network representations for such 2D fermionic SPTs.

Third, our fermionic Decorated Domain Wall procedure has a wide range of applications. Besides regular SPTs, one could apply these unitaries to produce fermionic versions of Floquet SPTs\cite{PotterVishwanathFidkowski2018,KumarDumitrescuPotter2018}, gapless SPTs\cite{Jiangetal2018,ScaffidiParkerVasseur2017,ParkerScaffidiVasseur2018} or subsystem SPTs\cite{Youetal2018}. The decorated domain wall procedure introduced could also possibly be extended to parafermions\cite{SonAlicea2018}.

Looking forward, a key question is whether one can produce symmetric FDLUs to realize (non-chiral) fermionic topological phases of interest such as topological insulators and topological superconductors protected by time reversal symmetry in 2+1D. These feature symmetry groups outside  supercohomology, and hence either an explicit construction or the identification of an obstruction would be of great interest and deepen our understanding of fermionic SPTs.

\begin{acknowledgements}
We would like to thank Iris Cong, Ruihua Fan, Chenjie Wang, and Yi-Zhuang You for stimulating discussions, Davide Gaiotto, Nicolas Tarantino, Alex Turzillo, and Beni Yoshida for correspondence, Tyler Ellison and Lukasz Fidkowski for sharing their unpublished preprint, and the referee for asking us to hone our definitions and arguments. N.T. is supported by the Purcell fellowship. A.V. is supported by a Simons Investigator grant.
\end{acknowledgements}

\appendix

\section{Decorated Domain Walls for Bosons}
\label{DDW}
\subsection{The Cluster State}
\label{cluster}
The cluster state, is a well-known 1D SPT protected by $G=\mathbb Z_2 \times \mathbb Z_2$ used predominantly in measurement-based quantum computation\cite{Sonetal2012,Yoshida2015,Yoshida2016,ZengZhou2016}. The Hamiltonian is given on a ring with $2N$ sites by
\begin{equation}
H = -\sum_{i=1}^{2N} \frac{1}{2}(1+Z_{i-1}X_iZ_{i+1}).
\end{equation}
The two symmetries are represented by flipping the spins on the odd and even sites respectively
\begin{align}
\chi_1 &= \prod_{i=1}^{N} X_{2i-1} &\chi_2 &= \prod_{i=1}^{N} X_{2i}.
\end{align}
One can obtain this Hamiltonian starting from the bare Hamiltonian,
\begin{equation}
H_0 = -\sum_{i=1}^{2N} \frac{1}{2}(1+X_i)
\end{equation}
and conjugating with the following overall symmetric FDLU
\begin{equation}
U = \prod_{i=1}^{2N} CZ_{i,i+1}.
\end{equation}
The action of this unitary on the product state realizes the decorated domain-wall procedure \cite{ChenLuVishwanath2014} by decorating a non-trivial 0D SPT of the second symmetry whenever there is a domain wall in the first symmetry. To see this, recall that there are two 0D SPTs with $\mathbb Z_2$ symmetry represented by $X$: $\ket{+}$ and $\ket{-}$. They can be transformed into each other via $Z$. In our evolution operator, the $CZ$ gate acts $Z$ on an even site whenever an adjacent odd site is $\ket{1}$. Thus, we see that a non-trivial 0D SPT on the even site gets created only when only one of the adjacent sites on the odd sites is $\ket{1}$, i.e., when the odd sites form a domain wall.

There is a similar Hamiltonian that is in the same phase as the cluster state. It is given by negating the signs of all the $X$ operators,
\begin{equation}
H' = -\sum_{i=1}^{2N} \frac{1}{2}(1-Z_{i-1}X_iZ_{i+1}).
\end{equation}
This can be accomplished by instead applying the unitary
\begin{equation}
U' = \prod_{i=1}^{2N} Z_i CZ_{i,i+1}.
\end{equation}
Since $\prod_{i} Z_i$ is a symmetric FDLU, the two ground states are in the same phase. The ground state of this wavefunction gives a minus sign to every spin down and every adjacent spin downs. In other words, the total sign is equal to the number of spin down regions, which equals the number of spin up regions, and so the ground state is $\mathbb Z_2$ symmetric.

It turns out that the cluster Hamiltonian and its cousin also realizes a $\mathbb Z_2^T$ SPT if time reversal symmetry is defined as
\begin{equation}
T = \prod_i X_i K,
\end{equation}
where $K$ is complex conjugation\cite{Santos2015}. One can check that $U$ and $U'$ are also overall symmetric with respect to $T$, and the edges of the corresponding dressed Hamiltonians have $T^2=-1$ projective representations.

\subsection{Decorated Domain Walls for $\mathbb Z_m \times \mathbb Z_m$ SPTs}
\label{ZN}
The Decorated Domain Wall construction naturally generalizes to $\mathbb Z_m \times \mathbb Z_m$ symmetry by changing qubits to qudits $\ket{h}$ where $h=0,...,m-1$ is defined modulo $m$. The Pauli $Z$ and $X$ are generalized to clock and shift matrices $\mathcal{Z}$ and $\mathcal{X}$ defined via their action
\begin{align}
\mathcal{Z} \ket{h} &= \varpi^h \ket{h},  \\
\mathcal{X} \ket{h} &= \ket{h+1},
\end{align}
where $\varpi = e^{2 \pi i/m}$. They satisfy
\begin{align}
\mathcal Z^m =\mathcal X^m &=1, &  \mathcal Z \mathcal X &= \varpi \mathcal X \mathcal Z.
\end{align}
Note that $\mathcal Z$ and $\mathcal X$ are no longer Hermitian. The bare Hamiltonian can be written using the projector at each site,
\begin{equation}
H_0 = -\sum_{i=1}^{2N} \ket{0}_i\bra{0} = -\sum_{i=1}^{2N} \frac{1}{m} \sum_{n=0}^{m-1} \mathcal X^n,
\end{equation}
which has symmetries generated by
\begin{align}
\chi_1 &= \prod_{i=1}^N \mathcal X_{2i-1}, & \chi_2 &= \prod_{i=1}^N \mathcal X_{2i}.
\end{align}

We evolve with the unitary operator
\begin{equation}
U = \prod_{i=1}^N C\mathcal Z^\dagger_{2i-1,2i} C \mathcal Z_{2i,2i+1},
\end{equation}
where the controlled-$\mathcal Z$ is defined to act as
\begin{equation}
C \mathcal Z_{i,j} \ket{h_i,h_j} = \varpi^{h_ih_j} \ket{h_i,h_j}.
\end{equation}
$U$ decorates the $\chi_1$ domain wall labeled by $h_i -h_{i-2}$ with the corresponding $\chi_2$ charge.

Conjugating, we can write the Hamiltonian in term of dressed projectors
\begin{equation}
\Pi_i = \frac {1}{m} \sum_{n=0}^{m-1}  \bar{\mathcal X}^n,
\label{equ:dressedZN}
\end{equation}
where
\begin{align}
\bar{\mathcal X}_{2i-1} &= \mathcal Z_{2i-2} \mathcal X_{2i-1} \mathcal Z_{2i}^\dagger, & \bar{\mathcal X}_{2i-1} &= \mathcal  Z_{2i-1}^\dagger \mathcal  X_{2i} \mathcal  Z_{2i+1}.
\end{align}

To obtain the projective representation of the edge, we set the dressed projectors to one when acting on the ground state. Multiplying Eq. \eqref{equ:dressedZN} by $\bar{\mathcal{X}} -1$ on both sides, we find that the left hand side is $\bar{\mathcal{X}}^m-1$, which is zero. Hence, $\bar{\mathcal{X}} =1$ on the ground state. From this, one can read off the projective representations on the left edge.
\begin{align}
\chi_1^L &= \mathcal X_1 \mathcal Z_2, &\chi_2^L &= \mathcal Z_1^{-1},
\end{align}
and we conclude that
\begin{equation}
\chi_1^L \chi_2^L = \varpi^{-1} \chi_2^L \chi_1^L.
\end{equation}
This is the projective representation of the generating phase.

We remark that similar SPT Hamiltonians have been previously constructed in Refs. \onlinecite{GeraedtsMotrunich2014,Tsuietal2017,KumarDumitrescuPotter2018}, but the Hamiltonians given are not commuting projectors. For a generalization to $\mathbb Z_m \times \mathbb Z_m \times \mathbb Z_m$ SPTs in 2D, see Refs. \onlinecite{Yoshida2016,ChenPrakashWei2018}.

\section{The Levin-Gu SPT}
\label{LevinGu}
The Levin-Gu model\cite{LevinGu2012} is a 2D bosonic SPT protected by $G=\mathbb Z_2$. It is given on a triangular lattice by the commuting projector Hamiltonian
\begin{widetext}
\begin{align}
H= -\sum_i \frac{1}{2} \left [ 1- X_0 i^{(1-Z_1Z_2)/2} i^{(1-Z_2Z_3)/2}i^{(1-Z_3Z_4)/2} i^{(1-Z_4Z_5)/2} i^{(1-Z_5Z_6)/2} i^{(1-Z_6Z_1)/2} \right].
\end{align}
where the indices are ordered according to Figure \ref{fig:ordering}. Writing $Z_i = 1-2g_i$, and expanding, we find that
\begin{align}
H= -\sum_i \frac{1}{2} \left [ 1- X_0 Z_1Z_2Z_3Z_4Z_5Z_6  CZ_{12} CZ_{23} CZ_{34} CZ_{45}CZ_{56} CZ_{61} \right].
\end{align}
\end{widetext}
Starting from the bare projector Hamiltonian $H_0 = -\sum_i \frac{1}{2}(1+X_i)$, we can evolve it to $H$ by the unitary
\begin{equation}
U= \prod_i Z_i \prod_{<ij>} CZ_{ij} \prod_{\Delta,\nabla} CCZ_{ijk}.
\end{equation}
The ground state wavefunction is $U\ket{\psi_0}$, where $\ket{\psi_0}$ is an equal superposition of all up and down states. In summary, $U$ gives a minus sign to every vertex, edge, and face that contains all spin downs. Hence, for a region of spin-downs, we get a factor of $(-1)^{V-E+F}=(-1)^{\chi_\downarrow}$, where $\chi_\downarrow$ is the total Euler characteristic of the spin-down regions. Note that since the Euler characteristic of a closed surface $\chi_{\mathcal M}$ is always even, and 
\begin{equation}
\chi_\downarrow +\chi_\uparrow = \chi_{\mathcal M},
\end{equation}
the sign factor from spin-up and spin-down regions are equal
\begin{equation}
(-1)^{\chi_\downarrow} = (-1)^{\chi_\uparrow},
\end{equation}
which reflects the $\mathbb Z_2$ symmetry of the wavefunction. For example, we get a minus sign for every closed disk of down spins, and we get an extra minus sign for each region of spin ups we build inside. Thus we can also find the total minus sign from the number of domain wall loops.

Note that in the unitary, $\prod_i Z_i \prod_{<ij>} CZ_{ij}$ is a symmetric FDLU on a triangular lattice, so in fact only $\prod_{\Delta,\nabla} CCZ_{ijk}$ was required to get the SPT phase\cite{Yoshida2015}. However, defining our unitary this way, we see that the Levin-Gu SPT can be defined on any arbitrary triangulation of the space manifold.

\section{Computing the Doubly Dressed Projectors in 1D}
\label{1Dapp}
In this appendix, we explicitly compute $U^2 X_i (U^2)^\dagger$ for the 1D SPT protected by $\mathbb Z_2^T \times \mathbb Z_2^F$, where $U^2$ is given by Eq. \eqref{equ:1DsignU2}. Since our system is on a ring, we can always relabel $i$ to 1 without loss of generality. We can divide up the sum in the exponent into three parts:\\
1. Terms that do not contain $g_1$ at all.\\
2. Terms that contain $g_1$ in one bracket.\\
3. Terms that contain $g_1$ in two brackets.\\
The first term does not contribute to $U^2X_1U^{\dagger2}$. The second term can be written out explicitly as
\begin{align}
(g_1+g_2) \left (\sum_{k=2}^{N-1} (g_k+g_{k+1}) \right)+ \left (\sum_{j=2}^{N-1}(g_j+g_{j+1})\right) (g_N+g_1) \nonumber\\
=(g_1+g_2)(g_2+g_N) + (g_2+g_N)(g_N+g_1) = g_2+g_N,
\end{align}
and we see that it actually does not depend on $g_1$. Thus the only term that contributes to the dressed qubit is the third term, which is
\begin{equation}
(g_1+g_2)(g_N+g_1).
\end{equation}
Therefore,
\begin{align}
U^2X_1U^{\dagger2} &= X_1 (-1)^{((1-g_1)+g_2)(g_N+(1-g_1))} (-1)^{(g_1+g_2)(g_N+g_1)} \nonumber\\
&= X_1 (-1)^{1+g_2+g_N} = -Z_N X_1 Z_2.
\end{align}
Shifting back to general $i$, we conclude that
\begin{equation}
U^2X_iU^{\dagger2} = -Z_{i-1}X_iZ_{i+1}.
\end{equation}

\section{Checking Properties of the 2D model}
\label{properties}
In this Appendix, we check that our Hamiltonian for $\nu=2$ in Eq. \eqref{equ:H_2} satisfies all the properties of a symmetric commuting projector. Properties \ref{Projector} and \ref{Hermitian} are automically satisfied by construction, and so are simply consistency checks. Properties \ref{SymmetricProjector}, \ref{SymmetricU}, must be explicitly checked in order for our model to be a valid SPT, and property \ref{U2givesnu4} proves that we have constructed the square root of the Levin-Gu phase.

\subsection{All Terms are Projectors}
\label{Projector}
It suffices to show that $\Pi_i =\frac{1}{2}(1+A_i)$ is a projector if $A_i^2=1$. Thus, the dressed fermion term
\begin{equation}
(-1)^{n_{i+\frac{1}{2}}}Z_iZ_jZ_kCZ_{ij}CZ_{jk}CZ_{ik}
\end{equation}
clearly gives a projector. For the dressed qubit term, first we find that squaring the Majorana terms give
\begin{align}
&(\gamma_{12}^{g_1 +g_2+1}  \gamma_{23}^{g_2 +g_3+1} \gamma_{34}^{g_3 +g_4+1}  \gamma_{45}^{g_4 +g_5+1} \gamma_{56}^{g_5 +g_6+1}  \gamma_{61}^{g_6 +g_1+1})^2 \nonumber\\
&= (-1)^{1+g_1+g_2+g_3+g_4+g_5+g_6+g_1g_2+g_2g_3+g_3g_4+g_4g_5+g_5g_6+g_6g_1} \nonumber\\
&=-Z_1 Z_2 Z_3 Z_4 Z_5 Z_6 CZ_{01}  CZ_{02}  CZ_{03}  CZ_{04}  CZ_{05}  CZ_{06}.
\label{equ:majoranasign}
\end{align}
Next, we see that since there is only one $X$ operator at site $0$, all qubit gates that don't involve the index $0$ square to one. The terms that involve $0$ must commute past $X_0$ and give sign factors. First, $Z_0$ anticommutes and gives the minus sign. Second, the six $CZ$'s in the second line of Eq. \eqref{equ:H_2} ($CZ_{0i}$ where $i=1,...,6$) give $X_0 CZ_{0i} X_0 CZ_{0i} = Z_i$. Lastly, squaring the three $C\Ri$ and $C\Ric$ terms gives $CZ$ around the hexagon. All these cancel the terms left over from squaring the Majoranas so the dressed $X$ squares to one.

\subsection{The Dressed Qubit Projector is Hermitian}
\label{Hermitian}
This is very similar to the previous proof. Upon conjugating, the order of the Majoranas is reversed and reordering them back gives a sign factor in Eq. \eqref{equ:majoranasign}. The minus sign cancels with anticommuting $Z_0$ and $X_0$. The product of $Z$'s around the hexagon come from swapping $X_0$ and $CZ_{0i}$, and the product of $CZ$'s come from conjugating $C\Ri$ and $C\Ric$.

\subsection{The Projectors Commute with the Symmetries}
\label{SymmetricProjector}
We check that each term commutes with the $\mathbb Z_2$ symmetry $\chi = \prod_i X_i$. It is clear that the dressed fermion commutes with $\chi$. For the dressed qubit, the controlled-Majorana operators remain invariant under $g_i \rightarrow 1-g_i$ and so they all commute with the symmetry. Next, the second line of Eq. \eqref{equ:H_2} has an odd number of $CZ$ operators forming a closed loop, so they anticommute with $\chi$, which cancels the anticommutation with $Z_0$. Finally, for the remaining $C\Ri$ and $C\Ric$ operators, using
\begin{align}
X_iX_j C\Ri_{ij} X_i X_j &=  i \Ric_i\Ric_jC\Ri_{ij},\\
X_iX_j C\Ric_{ij} X_i X_j &=  -i \Ri_i\Ri_jC\Ri_{ij},
\end{align}
we find that the product of the six terms remain invariant under conjugation by $\chi$.

To see that the projectors commute with fermion parity, we see that the total number Majoranas in each projector is always even.

\subsection{$U$ is an Overall Symmetric FDLU}
\label{SymmetricU}
Consider $U$ given by Eq. \eqref{equ:U2}. By construction, each gate in $U$ commutes with $P_f$, but does not commute with $\chi$. Nevertheless, we can still show that $U$ is overall symmetric by using the fact that the dressed qubit projectors in $H_2$ are symmetric.\\
Consider $\chi U \chi$ and first focus on the $CC\Ri$ and $CC\Ric$ gates. Upon conjugation, we find 
\begin{align}
 \chi C\Ri_{ijk} \chi&= i \Ric_i\Ric_j \Ric_k C\Ri_{ij} C\Ri_{ik} C\Ri_{jk} CCZ_{ijk} C\Ri_{ijk}, \\
 \chi C\Ric_{ijk}\chi &= -i \Ri_i\Ri_j \Ri_k C\Ric_{ij} C\Ric_{ik} C\Ric_{jk} CCZ_{ijk} C\Ric_{ijk}.
\end{align}
We see that $i$ and $\Ri$,$\Ric$,$C\Ri$,$C\Ric$ cancel upon multiplying over a closed manifold. Thus we only have signs from $CCZ_{ijk}$ from all triangles.\\
Next consider the controlled-Majorana gates. The physical site $\gamma_{(ijk)}^{g_ig_j+g_jg_k+g_ig_k+g_i+g_j+g_k}$ is invariant under $\chi$. The ancilla fermions are not, but since they always appear in pairs in the unitary, the only change must be a sign factor from the possible anticommutation of the Majoranas. Thus, we conclude that by conjugating $U$ with the symmetry, the most general form it can take is a sign
\begin{equation}
 \chi U \chi = U (-1)^{f(\{g_i\})}
 \label{equ:generalform}
 \end{equation}
 for some function $f(\{g_i\})$ which can possibly depend on the spin of all sites ${g_i}$.\\
Now fix a qubit site $0$. We have shown that the dressed operator $UX_IU^\dagger$ is invariant under the symmetry. Thus,
\begin{equation}
 U X_0 U^\dagger = \chi U X_0 U^\dagger \chi =  (\chi U \chi) X_0 (\chi U^\dagger \chi).
  \end{equation}
Substituting the general form from Eq. \eqref{equ:generalform}, we obtain
\begin{equation}
 U X_0 U^\dagger = U (-1)^{f(\{g_i\})} X_0 (-1)^{f(\{g_i\})} U^\dagger.
  \end{equation}
Therefore, we find that modulo 2, $f$ needs to satisfy
\begin{equation}
f(\{g_i\}) = \left.f(\{g_i\}) \right|_{g_0 \rightarrow 1-g_0}
\end{equation}
for all qubit sites. Let us write down the most general possible form of $f$ 
\begin{equation}
f = c+ \sum_i c_i g_i + \sum_{i<j} c_{ij} g_i g_j + \sum_{i<j<k} c_{ijk} g_ig_jg_k + \cdots,
\end{equation}
where $c,c_i,c_{ij},...=0,1$. Under $g_0 \rightarrow 1-g_0$, the difference must be zero 
\begin{equation}
f - f|_{g_0 \rightarrow 1-g_0} =  c_0 + \sum_{j \ne 0} c_{0j} g_j + \sum_{\substack{j<k\\j,k\ne 0}} c_{0jk} g_jg_k =0.
\end{equation}
Hence, we see that in order for the equation to be true for all $j,k,... \ne 0$ all the coefficients above must be zero. Repeating for all qubit sites, we conclude that only $c$ can be non-zero. Thus we have shown that $U$ commutes with the $\mathbb Z_2$ symmetry $\chi$ up to a sign $(-1)^c$, which amounts to our choice of spin structure on the space manifold. Thus, $U$ is an overall symmetric FDLU.

\subsection{$U^2$ Gives the Levin-Gu Hamiltonian}
\label{U2givesnu4}
We will show that evolving the bare Hamiltonian with $U^2$ gives exactly the Levin-Gu model, which is the $\nu=4$ phase.
Similarly to the 1D case, upon squaring we find that all the Majorana operators have squared away
\begin{equation}
U^2 =\prod_{\Delta,\nabla} CCZ (-1)^{F(\{g_i\})}
\end{equation}
leaving a sign $(-1)^{F(\{g_i\})}$ from possible anticommutations. We can forget about the ancilla qubits because pairing them up in $U$ produces a sign factor, which disappears upon squaring. Hence, we only have to consider the physical fermions at the center of the triangles. As a shorthand, define
\begin{equation}
s_{(ijk)} = g_i+g_j+g_k + g_ig_j + g_jg_k + g_ig_k,
\end{equation}
then we can explicitly write down the sign factor as
\begin{equation}
F(\{g_i\}) = \sum_{\Delta_{(ijk)} < \Delta_{(i'j'k')}} s_{(ijk)} s_{(i'j'k')},
\end{equation}
where we imposed a certain ordering for our triangles.

 Since $U^2$ leaves the fermions unchanged, we only need to conjugate $X_0$ with $U^2$, which gives
\begin{align}
U^2 X_0 U^{2\dagger}=& X_0 CZ_{12} CZ_{23} CZ_{34} CZ_{45} CZ_{56} CZ_{61} \nonumber\\
&\times \left. (-1)^{F(\{g_i\})} \right|_{g_0 \rightarrow 1-g_0} (-1)^{F(\{g_i\})}.
\label{equ:signU2}
\end{align}
Now, we would like to find the contribution of $F(\{g_i\})$ to the equation above. Let us break $F({g_i})$ up into three terms:\\
1. $F_1$ contains terms that do not contain $g_0$ at all.\\
2. $F_2$ contains terms that contain $g_0$ in one triangle.\\
3. $F_3$ contains terms that contain $g_0$ in two triangles.\\
The first part will not contribute to Eq. \eqref{equ:signU2}, and we can show that $F_2$ also does not contribute as follows. With Figure \ref{fig:ordering} in mind, consider a term containing the vertex $0$, $(0jk)$. Without loss of generality, we can order the triangles so that this triangle is first, then the second contribution contains
\begin{equation}
s_{(0jk)} \left ( \sum_{\Delta\text{ not touching 0}} s_{(i'j'k')} \right )
\end{equation}
On a closed manifold, the sum above reduces to terms only on the boundary of the hexagon 
\begin{equation}
g_1+g_2+g_3+g_4+g_5+g_6 + g_1g_2+g_2g_3+g_3g_4+g_4g_5+g_5g_6+g_6g_1.
\end{equation}
Now, when we consider the contribution to Eq. \eqref{equ:signU2}, $s_{(0jk)}$ changes by to $(1+g_j+g_k)$ under $g_0 \rightarrow 1-g_0$. However, there are six terms like these around the hexagon multiplying the same sum. Thus, the sum of the six terms cancels out and $F_2$ does not contribute.

We are left with only $F_3$ which contains products of all pair combinations of the six triangles containing site $0$. Computing explicitly, we find that
\begin{equation}
\left. (-1)^{F_3}\right|_{g_0 \rightarrow 1-g_0} (-1)^{F_3} = (-1)^{1+g_1+g_2+g_3+g_4+g_5+g_6}.
\end{equation}
 Our resulting Hamiltonian is therefore
 \begin{widetext}
\begin{equation}
H_4= -\sum_i\frac{1}{2}(1- X_0 Z_1Z_2Z_3Z_4Z_5Z_6 CZ_{12} CZ_{23} CZ_{34} CZ_{45}CZ_{56} CZ_{61}) -\sum_{\Delta,\nabla}\frac{1}{2}(1+(-1)^{n_{(ijk)}}),
\end{equation}
\end{widetext}
which is exactly the Levin-Gu Hamiltonian with bare projectors of fermions.

\section{Topological Invariant for the 1D $\mathbb Z_2 \times \mathbb Z_2^F$ SPT}
\label{topinv}
$\mathbb Z_2 \times \mathbb Z_2^F$ SPTs are classified by $\mathbb Z_2 \times \mathbb Z_2$. One of the generators is a $\mathbb Z_2$ invariant Majorana chain. The other generator is a supercohomology phase. In analyzing the edge of our 2D model, we claim that the edge Hamiltonian \eqref{equ:edge} has the same critical properties as a critical point between a trivial SPT and the latter phase.

The non-trivialness of the supercohomology phase can be detected by a topological invariant identified in Ref. \onlinecite{Tantivasadakarn2017}. In Euclidean space-time, the invariant is a quotient of two partition functions on $T^2$: one with $\mathbb  Z_2$ and $\mathbb Z_2^F$ fluxes inserted into each cycle, and one with only the $\mathbb Z_2$ flux. In Minkowski space, they correspond to taking a trace of the $\mathbb Z_2$ symmetry $\chi = \prod_i X_i$ times the density matrix with R/NS spin structures respectively. Hence the invariant can be written as
\begin{equation}
\frac{_R\bra{\psi} \chi\ket{\psi}_{R}}{_{NS}\bra{\psi} \chi\ket{\psi}_{NS}}.
\end{equation}
This invariant is 1 for the trivial phase and -1 for the supercohomology phase.

As an example, it turns out that Eq. \eqref{equ:H21D} is a Hamiltonian that realizes this supercohomology phase if complex conjugation is removed from $\mathbb Z_2^T$. This is because all the operators used were real, and so removing complex conjugation does not change any of the derivations in section \ref{1Dmodel}. Since we have established that $U$ anticommutes/commutes with the $\mathbb Z_2$ symmetry when the fermions have periodic/antiperiodic boundary conditions, respectively; $_R\bra{\psi} \chi\ket{\psi}_{R}$ and $_{NS}\bra{\psi} \chi\ket{\psi}_{NS}$ must have opposite sign, which shows that this Hamiltonian realizes the supercohomology phase.

\bibliography{references}

\end{document}